\documentclass[11pt,a4paper]{article}
\pdfoutput=1

\usepackage{amsmath}
\usepackage[T1]{fontenc}

\usepackage{jheppub}
\usepackage{psfrag}
\usepackage{slashed}
\usepackage{cancel}
\usepackage{lscape}

\usepackage{caption}
\usepackage{array}
\usepackage{graphicx}
\usepackage{subcaption}
\usepackage{multirow}
\usepackage{tabularx}
\usepackage{makecell}

\usepackage[utf8]{inputenc}

\renewcommand{\arraystretch}{1.3}

\usepackage[utf8]{inputenc}
\usepackage{graphicx}
\usepackage{slashed}
\usepackage{color}
\usepackage{amsmath}
\usepackage{amssymb}

\definecolor{mypink}{RGB}{219, 48, 122}
\definecolor{mygreen}{rgb}{0,0.7,0}

\def\nn{\nonumber \\ }

\def\la{\langle}
\def\ra{\rangle}
\def\spA#1#2{\la#1#2\ra}
\def\spB#1#2{[#1#2]}
\def\spAB#1#2#3{\la#1|#2|#3]}

\def\spAA#1#2#3{\la#1|#2|#3\ra}
\def\spBB#1#2#3{[#1|#2|#3]}

\DeclareMathOperator{\tr}{\rm tr}
\def\trm{\tr_-}
\def\trp{\tr_+}

\def\eps{\epsilon}

\def\bbh{b\bar{b}H}
\def\bbggh{\bar{b}bggH}
\def\bbqqh{\bar{b}b\bar{q}qH}

\def\cusp{{\mathrm{cusp}}}
\def\sumhel{\sum_{\mathrm{helicity}}}

\def\hpl11{{\mathrm{HPL}}_{1,1}}

\newcolumntype{C}[1]{>{\hsize=#1\hsize\centering\arraybackslash}X}%

\newcolumntype{Z}{r<{\hspace{3mm}}}

\newcommand{\ba}{\[\begin{aligned}}
\newcommand{\ea}{\end{aligned}\]}

\title{Two-loop leading-colour QCD helicity amplitudes for Higgs boson production in association with a bottom-quark pair at the LHC}

\author[a]{Simon Badger,}
\author[b]{Heribertus Bayu Hartanto,}
\author[a,c]{Jakub Kry\'s,}
\author[a]{Simone Zoia}

\affiliation[a]{
Dipartimento di Fisica and Arnold-Regge Center, Università di Torino, and INFN, Sezione di
Torino, Via P. Giuria 1, I-10125 Torino, Italy
}

\affiliation[b]{
Cavendish Laboratory, University of Cambridge, Cambridge CB3 0HE, United Kingdom
}

\affiliation[c]{
Institute for Particle Physics Phenomenology, Department of Physics, Durham University, Durham DH1 3LE, United Kingdom
}

\emailAdd{
simondavid.badger@unito,it,
hbhartanto@hep.phy.cam.ac.uk,
jakubmarcin.krys@unito.it,
simone.zoia@unito.it
}

\abstract{
We compute the two-loop QCD helicity amplitudes for the production of a Higgs boson in association
with a bottom quark pair at a hadron collider. We take the approximations of leading colour and work in the five flavour scheme, where the bottom quarks are massless while the Yukawa coupling is non-zero. 
We extract analytic expressions from multiple numerical
evaluations over finite fields and present the results in terms of an independent set of special
functions that can be reliably evaluated over the full phase space.
}

\preprint{CAVENDISH-HEP-21/11}

\begin{document}
\maketitle
\flushbottom
\section{Introduction \label{sec:intro}}

Precise theoretical predictions are an essential ingredient in the search for subtle deviations from
the Standard Model (SM) at current and future collider experiments. The enormous amount of data
gathered at the Large Hadron Collider (LHC) is already challenging the theoretical precision for
many processes and the pressure will increase as data continues to pour in. It has been clear for
some time~\cite{Badger:2016bpw,Bendavid:2018nar,Amoroso:2020lgh} that, for a large class of two- and three-particle final states, at least the
Next-to-Next-to-Leading Order corrections in Quantum Chromodynamics (NNLO QCD) will be required for
fully differential cross sections. The high multiplicity final states, particularly those with many kinematic
scales, pose the biggest technical challenge owing to the enormous analytic and algebraic
complexity. Nevertheless, thanks to a huge effort across the theoretical community, new tools and
methods have been developed that have produced the first NNLO QCD predictions for $2\to3$
processes~\cite{Chawdhry:2019bji,Kallweit:2020gcp,Chawdhry:2021hkp}, most notably for 3-jet
production~\cite{Czakon:2021mjy}. This remarkable progress has been driven both by the advancements
in the scattering amplitude computation and by efficient subtraction schemes. By now, all the
two-loop master integrals --~which are one of the important ingredients in the amplitude
computation~-- are available for the fully massless five-particle
processes~\cite{Papadopoulos:2015jft,Gehrmann:2018yef,Abreu:2018rcw,Chicherin:2018mue,Chicherin:2018old,Chicherin:2020oor},
allowing for several two-loop QCD amplitudes to be derived
analytically~\cite{Gehrmann:2015bfy,Badger:2018enw,Abreu:2018zmy,Abreu:2019odu,Badger:2019djh,Abreu:2020cwb,Chawdhry:2020for,Agarwal:2021grm,Abreu:2021fuk,Chawdhry:2021mkw,Agarwal:2021vdh,Badger:2021imn}, improving on previous results that were obtained
numerically~\cite{Badger:2017jhb,Abreu:2017hqn,Badger:2018gip,Abreu:2018jgq}. These new results have
been achieved thanks to technological breakthroughs in the method of differential
equations~\cite{Kotikov:1990kg,Bern:1993kr,Remiddi:1997ny,Gehrmann:1999as,Henn:2013pwa}, integral
reduction algorithms~\cite{Gluza:2010ws,Schabinger:2011dz,Ita:2015tya,Larsen:2015ped,Boehm:2017wjc}
and the use of finite-field arithmetic to tame the algebraic complexity of multi-leg and multi-scale
problems~\cite{vonManteuffel:2014ixa,Peraro:2016wsq,Klappert:2019emp,Peraro:2019svx,Klappert:2020aqs,Klappert:2020nbg}.

A different class of $2\to 3$ processes that is of great interest involves an external massive leg.
The two-loop planar helicity amplitudes for $W+4$-parton scattering (contributing to the prediction
for $W+2$-jet production at NNLO QCD) have been previously studied numerically~\cite{Hartanto:2019uvl}. 
All two-loop integrals needed for this type of process are available for the planar
contributions~\cite{Papadopoulos:2015jft,Abreu:2020jxa,Canko:2020ylt,Syrrakos:2020kba}, and the
first analytic result was derived for $Wb\bar{b}$ production in the leading colour, massless
$b$-quark and on-shell $W$-boson approximations~\cite{Badger:2021nhg}. 
Very recently, complete analytic results for one of the non-planar integral topologies have also become available~\cite{Papadopoulos:2019iam,abreu2021twoloop}.

In this article we consider the two-loop amplitudes relevant for the production of a Higgs boson in
association with a bottom-quark pair, i.e. $pp \to b\bar{b}H$, in the leading colour approximation.
$b\bar{b}H$ production at the LHC has been a subject of great phenomenological interest due to its
potential in directly measuring the bottom-quark Yukawa coupling. In the Standard Model (SM), the
coupling strengths of the Higgs boson to the fermions and vector bosons are proportional to their
mass, causing the rate of the $b\bar{b}H$ production to be suppressed with respect to, for example,
Higgs production in gluon fusion ($gg\to H$) or vector boson fusion ($pp\to Hjj$), associated
production with a vector boson ($pp\to VH$), and associated production with a top-quark pair ($pp\to
t\bar{t}H$). In addition, the presence of $b$-tagging further suppresses the $b\bar{b}H$ production
rate.  In some new physics scenarios, such as the Two Higgs Doublet Models (2HDM's) and the Minimal
Supersymmetric Standard Model (MSSM), the bottom-quark Yukawa coupling can be dramatically enhanced,
resulting in a considerable increase of the $b\bar{b}H$ production
rate~\cite{Balazs:1998nt,Dawson:2005vi}.  Thus, the study of the $b\bar{b}H$ production will allow
to constrain supersymmetric models and other extensions of the SM that modify the bottom-quark
Yukawa coupling. Recent studies on the interplay between $b\bar{b}H$ signal and backgrounds can be
found in Refs.~\cite{Pagani:2020rsg,Grojean:2020ech,Konar:2021nkk}.

The theoretical approach to obtaining predictions for the $pp\to b\bar{b}H$ process has been subject
of much discussion in the community. This is due to the fact that, in the presence of bottom quarks,
a theoretical prediction can be computed in either the four-flavour scheme (4FS) or the five-flavour
scheme (5FS). In the 4FS computation, bottom quarks are treated as massive and they do not
contribute to the parton distributions functions (PDFs), hence only appearing in the final state.
Large logarithms of the form $\log(m_b/Q)$ with $Q \propto m_H$ arise when the integration over the
bottom-quark phase space is performed, and such contributions may spoil the convergence of the
perturbative series. These large logarithms can be resummed to all orders by introducing the
bottom-quark parton distribution functions.  The 5FS approach stems from this prescription, where
the bottom-quarks are included in the parton distribution functions, allowed to appear in the
initial state, and treated as massless.  We refer the reader to Ref.~\cite{Maltoni:2012pa} for further
discussion on the 5FS and 4FS approaches. 
In 5FS the inclusive $b\bar{b}H$ production (where the tree level process is $b\bar{b}\to H$) has been computed up to
N3LO QCD~\cite{Dicus:1998hs,Balazs:1998sb,Maltoni:2003pn,Harlander:2010cz,Buehler:2012cu,Harlander:2012pb,H:2019nsw,Duhr:2019kwi,Mondini:2021nck}, 
while for the case where a single bottom quark is observed NLO QCD~\cite{Campbell:2002zm}, weak~\cite{Dawson:2010yz} 
and SUSY QCD~\cite{Dawson:2011pe} corrections are available.
In 4FS the $b\bar{b}H$ production has been calculated up to 
NLO QCD~\cite{Dittmaier:2003ej,Dawson:2003kb,Dawson:2004sh,Wiesemann:2014ioa,Jager:2015hka,Deutschmann:2018avk},
and the supersymmetric QCD corrections~\cite{Liu:2012qu,Dittmaier:2014sva} are also known.  
There have also been efforts in matching the 5FS and 4FS calculations to obtain accurate predictions across the entire kinematic region~\cite{Harlander:2011aa,Bonvini:2015pxa,Forte:2015hba,Forte:2016sja,Duhr:2020kzd}. 
A first step towards a massive version of the five-flavour scheme (5FMS) has been devised to naturally connect the 4FS and 5FS approaches~\cite{Krauss:2017wmx,Figueroa:2018chn}.

In this work we compute the two-loop QCD corrections to the $gg \rightarrow b\bar{b}H$,
$q\bar{q}/\bar{q}q \rightarrow b\bar{b}H$, $b\bar{b}/\bar{b}b\rightarrow b\bar{b}H$, $bb \rightarrow
bbH$ and $\bar{b}\bar{b}\rightarrow\bar{b}\bar{b}H$ reactions in the 5FS. These two-loop amplitudes
enter the computation of $pp(b\bar{b})\to H$ at N4LO, $pp\to b(\bar{b})H$ at N3LO when one $b$-jet
is tagged, and $pp\to b\bar{b}H$ at NNLO when two $b$-jets are required in the final state.  We note
that beyond NLO, for the computation with massless bottom quarks, a flavoured jet
algorithm~\cite{Banfi:2006hf} would have to be employed when identifying the $b$-jets, since the use
of conventional $k_T$ or anti-$k_T$ jet algorithms would render the fixed order computation infrared
unsafe.  We further remark that the two-loop amplitudes for $b\bar{b}H$ production derived in this
work can also be used in the computation of Higgs decaying into a bottom-quark pair in 5FS, by
crossing initial partons to the final state. Specifically, they will contribute in the N4LO $H\to
b\bar{b}$, N3LO $H\to b\bar{b}j$ and NNLO $H\to b\bar{b}jj$ computations.  In addition, by crossing
the $b\bar{b}$ pair to the initial state and the $gg/q\bar{q}$ pair to the final state we obtain the
contribution of the bottom-quark initiated channel to $H+2j$ production ($b\bar{b}\to Hjj$).

We present analytic results for the finite remainders after ultraviolet (UV) and infrared (IR) poles have been subtracted.
This is possible using a basis of independent special functions recently identified in the context
of $Wb\bar{b}$ production~\cite{Badger:2021nhg}. We obtain numerical results valid across the full phase
space by applying the generalised series expansion approach~\cite{Francesco:2019yqt,Abreu:2020jxa,Hidding:2020ytt} to
the differential equations satisfied by the special functions appearing in the finite remainders.

This paper is organised as follows. We begin by describing the structure of the $b\bar{b}H$
amplitudes at leading colour in Section~\ref{sec:amp}, followed by a discussion of the methodology
used in deriving the analytic expression of the amplitudes in Section~\ref{sec:reduction}. 
A description of the basis of special functions is presented in Section~\ref{sec:Hbasis} and a number of validations
that we performed on the results derived in this work are discussed in Section~\ref{sec:validation}.
We present benchmark numerical evaluations together with evaluations on a physical phase-space slice in
Section~\ref{sec:results}. Finally we draw our conclusions in Section~\ref{sec:conclusions}. 

\section{Structure of the $pp\to\bbh$ Amplitudes at Leading Colour}
\label{sec:amp}

We compute the two-loop QCD corrections in the leading colour approximation for the following subprocesses
\begin{align} \label{eq:subprocesses}
& 0 \rightarrow \bar{b}(p_1) + b(p_2) + g(p_3) + g(p_4) + H(p_5)\,, \\
& 0 \rightarrow \bar{b}(p_1) + b(p_2) + \bar{q}(p_3) + q(p_4) + H(p_5)\,, \\
& 0 \rightarrow \bar{b}(p_1) + b(p_2) + \bar{b}(p_3) + b(p_4) + H(p_5)\,,
\end{align}
where all momenta are taken as outgoing, 
\begin{align}
\sum_{i=1}^5 p_i = 0\,.
\end{align}
We work in the five-flavour scheme, where the bottom quark is taken as massless while its Yukawa coupling to the Higgs boson is kept finite, so that
\begin{align}
p_1^2=p_2^2=p_3^2=p_4^2=0\,, \qquad \quad p_5^2=m_H^2\,, 
\end{align}
where $m_H$ is the mass of the Higgs boson. The kinematics is described by six independent scalar products, which we choose as 
\begin{equation}
 (s_{12},s_{23},s_{34},s_{45},s_{15},p_5^2) \,, \nonumber
\end{equation} with $s_{ij} = (p_i + p_j)^2$, and a pseudo-scalar invariant,
\begin{align}
\text{tr}_5 = 4 i \epsilon_{\mu \nu \rho \sigma} p_1^{\mu} p_2^{\mu} p_3^{\mu} p_4^{\mu} = [12] \spA{2}{3} [34] \spA{4}{1} - \spA{1}{2} [23] \spA{3}{4} [41] \,.
\end{align}
The latter is connected to the Gram determinant of the four linearly independent momenta, $\Delta = \mathrm{det}(s_{ij})$ with $1 \leq i,j \leq 4$, via 
$\Delta = (\text{tr}_5)^2$.
 
The colour decomposition of the $L$-loop amplitudes in the leading colour approximation is given by
\begin{align}
&\mathcal{A}^{(L)}(1_{\bar b},2_{b},3_{g},4_{g} ,5_{H}) = n^L g_s^2 y_b \bigg[ 
  (T^{a_3} T^{a_4})_{i_2}^{\;\;\bar i_1} \,  A^{(L)}(1_{\bar b},2_{b},3_{g},4_{g} ,5_{H}) + ( 3 \leftrightarrow 4 ) \bigg]\,, \nonumber\\
&\mathcal{A}^{(L)}(1_{\bar b},2_{b},3_{\bar{q}},4_{q} ,5_{H}) = n^L  g_s^2 y_b 
 \delta_{i_4}^{\;\;\bar i_1} \delta_{i_2}^{\;\;\bar i_3}  \,
  A^{(L)}(1_{\bar b},2_{b},3_{\bar{q}},4_{q} ,5_{H}) \,,  \label{eq:colourdecomposition} \\
&\mathcal{A}^{(L)}(1_{\bar b},2_{b},3_{\bar{b}},4_{b} ,5_{H}) = n^L  g_s^2 y_b \bigg[
  \delta_{i_4}^{\;\;\bar i_1} \delta_{i_2}^{\;\;\bar i_3}  \, \biggl( A^{(L)}(1_{\bar b},2_{b},3_{\bar{q}},4_{q} ,5_{H}) + A^{(L)}(3_{\bar b},4_{b},1_{\bar{q}},2_{q} ,5_{H}) \biggr) \,  \nonumber\\
& \hspace{4.5cm} -\delta_{i_2}^{\;\;\bar i_1} \delta_{i_4}^{\;\;\bar i_3}  \, \biggl( A^{(L)}(1_{\bar b},4_{b},3_{\bar{q}},2_{q} ,5_{H}) + A^{(L)}(3_{\bar b},2_{b},1_{\bar{q}},4_{q} ,5_{H}) \biggr) \bigg]\,, 
\nonumber 
\end{align}
where $n= m_\eps  \alpha_s/(4\pi),\ \alpha_s = g_s^2/(4\pi)$, $m_\eps=i (4\pi)^{\eps} e^{-\eps\gamma_E}$, $T^a$ 
are the fundamental generators of $SU(N_c)$ normalised such that $\tr(T^a T^b) = \delta^{ab}$,
while $g_s$ and $y_b$ are the strong coupling constant and bottom-quark Yukawa coupling, respectively. 
We further decompose the partial amplitudes at one and two loops based on their closed fermion loop contributions, as
\begin{align}
& A^{(1)} = N_c A^{(1),1} + n_f A^{(1),n_f}  \,,  
\label{eq:NfDecomposition1L} \\
& A^{(2)} = N_c^2 A^{(2),1} + N_c n_f A^{(2),n_f} + n_f^2 A^{(2),n_f^2}  \,, 
\label{eq:NfDecomposition2L}
\end{align}
where $n_f$ is the number of light quarks circulating in the loop.
The Feynman diagrams with the Higgs boson directly coupled to a closed bottom-quark loop vanish since, for a massless bottom-quark, they contain a Dirac trace with an odd number of $\gamma$ matrices. 
In our computation we do not consider the closed top-quark loop contribution. 

\begin{figure}[t]
  \begin{center}
    \includegraphics[width=0.95\textwidth]{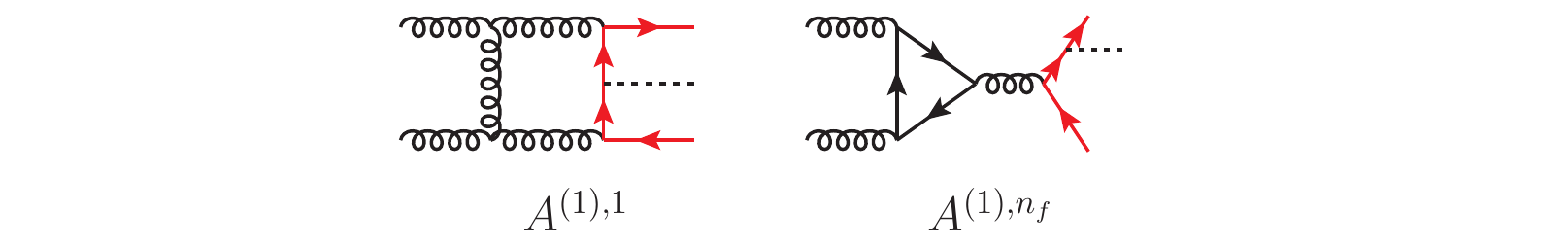}
  \end{center}
  \caption{Sample Feynman diagrams corresponding to the various closed fermion loop contributions at one loop as specified in Eq.~\eqref{eq:NfDecomposition1L}. 
  Black-dashed, red, black-spiralled and black lines represent Higgs bosons, bottom quarks, gluons and light quarks (bottom quarks included), respectively.}
  \label{fig:amp1L}
\end{figure}

\begin{figure}[t]
  \begin{center}
    \includegraphics[width=0.95\textwidth]{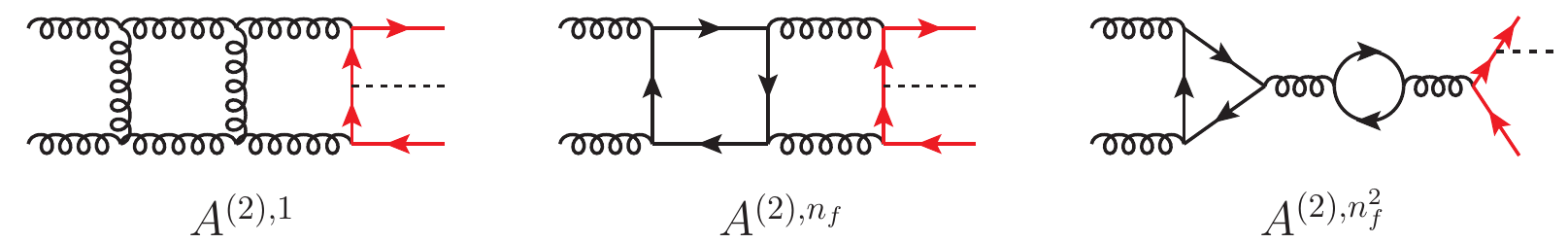}
  \end{center}
  \caption{Sample Feynman diagrams corresponding to the various closed fermion loop contributions at two loops as specified in Eq.~\eqref{eq:NfDecomposition2L}. 
  Black-dashed, red, black-spiralled and black lines represent Higgs bosons, bottom quarks, gluons and light quarks (bottom quarks included), respectively.}
  \label{fig:amp2L}
\end{figure}

The pole structure of the unrenormalised amplitudes in the 't Hooft-Veltman (tHV) scheme at one and two loops is given by
\begin{align}
P^{(1)} & = 2 I_1(\eps) + \frac{\beta_0}{\eps} - 2 s_1,  
\label{eq:poles1L} \\
P^{(2)} & =   2 I_1(\eps) \bigg( \hat{A}^{(1)} - \frac{\beta_0}{\eps} + 2 s_1\bigg) + 4 I_2(\eps)
                + \bigg(\frac{2\beta_0}{\eps}-2 s_1\bigg) \hat{A}^{(1)} \nn 
& \hspace{0.4cm} - \frac{\beta_0^2}{\eps^2} + \frac{\beta_1}{2\eps} - 4 s_2 + \frac{2 s_1\beta_0}{\eps} \,,
\label{eq:poles2L}
\end{align}
where $\hat{A}^{(1)}$ is the unrenormalised one-loop amplitude normalised to the tree-level amplitude.
$s_1$ and $s_2$ are the bottom-quark Yukawa renormalisation constants, and their expressions can be found in Appendix~\ref{app:renormconstants}.
We used a mixed renormalisation scheme where the strong coupling $\alpha_s$ and the bottom-Yukawa coupling $y_b$ are renormalised in the $\overline{\text{MS}}$ scheme, while the bottom-quark mass and wave function are renormalised in the on-shell ($\mathrm{OS}$) scheme. This allows to keep $y_b$ finite while taking the bottom-quark mass smoothly to zero ($m_b^{\mathrm{OS}}\rightarrow 0$)~\cite{Mondini:2021nck}.
Such a mixed renormalisation scheme can be used so long as pure QCD corrections are considered. In fact, using the $\overline{\text{MS}}$ scheme to renormalise $y_b$ allows us to better control the convergence of the perturbative corrections by resumming the large logarithms that appear in the OS scheme by running $y_b$ to a scale close to the Higgs mass. In the presence of electroweak (EW) corrections, however, the relationship between
$y_b$ and $m_b$ must be imposed to guarantee the cancellation of UV singularities~\cite{Pagani:2020rsg}.

The $I_2(\eps)$ operator is defined by
\begin{equation}
I_2(\eps) =   - \frac{1}{2}I_1(\eps) \left[ I_1(\eps)
              + \frac{\beta_0}{\eps} \right]
              + \frac{N(\eps)}{N(2\eps)} \left[ \frac{\beta_0}{2\eps}
                        + \frac{\gamma_{1}^{\cusp}}{8} \right] I_1(2\eps)
              + H^{(2)}(\eps) \,,
\end{equation}
while the $I_1(\eps)$ operators for $\bbh$ production in both the $gg$ and the $q\bar{q}$ channels are given at leading colour by
\begin{align}
I^{\bbqqh}_1(\eps) &= -N_c \frac{N(\eps)}{2} \bigg( \frac{1}{\eps^2} + \frac{3}{2\eps} \bigg) \big[ \left(-s_{23}\right)^{-\eps} + \left(-s_{14}\right)^{-\eps} \big], 
\label{eq:I1bbqqh}\\
I^{\bbggh}_1(\eps) &= -N_c \frac{N(\eps)}{2} \bigg\lbrace \bigg( \frac{1}{\eps^2} + \frac{3}{4\eps} + \frac{\beta_0}{4\eps} \bigg) \big[ \left(-s_{23}\right)^{-\eps} + \left(-s_{14}\right)^{-\eps} \big]
                 + \bigg( \frac{1}{\eps^2} + \frac{\beta_0}{2\eps} \bigg) \left(-s_{34}\right)^{-\eps} \bigg\rbrace \,,
\label{eq:I1bbggh}
\end{align}
where $N(\eps) = {e^{\eps \gamma_E}}/{\Gamma(1-\eps)}$, $s_{14} = s_{23}-s_{15}-s_{45}+p_5^2$ and
\begin{align}
H_{\bbqqh}^{(2)}(\eps) &= \frac{1}{16\eps} \bigg\lbrace 4 \gamma_1^q - \gamma_1^\cusp \gamma_0^q + \frac{\pi^2}{4} \beta_0 \gamma_0^\cusp C_F \bigg\rbrace \,, \\
H_{\bbggh}^{(2)}(\eps) &=  \frac{1}{16\eps} \bigg\lbrace 2\left(\gamma_1^q+\gamma_1^g\right)
                                                - \frac{1}{2}\gamma_1^\cusp \left(\gamma_0^q+\gamma_0^g\right)
                                                + \frac{\pi^2}{8} \beta_0 \gamma_0^\cusp \left(C_F+C_A\right) \bigg\rbrace \,.
\end{align}
The $\beta$ function coefficients and anomalous dimensions are given in Appendix~\ref{app:renormconstants}.
The finite remainder of the $L$-loop partial amplitude is then obtained by subtracting the poles $P^{(L)}$ 
(which include both the ultraviolet and infrared singularities) from the unrenormalised partial amplitude $A^{(L)}$ and setting $\eps$ to $0$,
\begin{equation}
F^{(L)} = \lim_{\eps \to 0} \left[ A^{(L)} - P^{(L)} A^{(0)} \right]\,.
\label{eq:finiteremainder}
\end{equation}
The partial finite remainders $F^{(L)}$ inherit from the partial amplitudes the decomposition in powers of $n_f$,
\begin{align}
& F^{(1)} = N_c F^{(1),1} + n_f F^{(1),n_f}  \,,  
\label{eq:NfDecomposition1LF} \\
& F^{(2)} = N_c^2 F^{(2),1} + N_c n_f F^{(2),n_f} + n_f^2 F^{(2),n_f^2}  \,.
\label{eq:NfDecomposition2LF}
\end{align}
The full finite remainders $\mathcal{F}^{(L)}$ are obtained from the partial ones $F^{(L)}$ through a colour decomposition analogous to that given in Eq.~\eqref{eq:colourdecomposition} for the bare amplitudes,
\begin{align}
&\mathcal{F}^{(L)}(1_{\bar b},2_{b},3_{g},4_{g} ,5_{H}) = n^L g_s^2 y_b \bigg[ 
  (T^{a_3} T^{a_4})_{i_2}^{\;\;\bar i_1} \,  F^{(L)}(1_{\bar b},2_{b},3_{g},4_{g} ,5_{H}) + ( 3 \leftrightarrow 4 ) \bigg]\,, \nonumber\\
&\mathcal{F}^{(L)}(1_{\bar b},2_{b},3_{\bar{q}},4_{q} ,5_{H}) = n^L  g_s^2 y_b 
 \delta_{i_4}^{\;\;\bar i_1} \delta_{i_2}^{\;\;\bar i_3}  \,
  F^{(L)}(1_{\bar b},2_{b},3_{\bar{q}},4_{q} ,5_{H}) \,,  \label{eq:colourdecompositionF} \\
&\mathcal{F}^{(L)}(1_{\bar b},2_{b},3_{\bar{b}},4_{b} ,5_{H}) = n^L  g_s^2 y_b \bigg[
  \delta_{i_4}^{\;\;\bar i_1} \delta_{i_2}^{\;\;\bar i_3}  \, \biggl( F^{(L)}(1_{\bar b},2_{b},3_{\bar{q}},4_{q} ,5_{H}) + F^{(L)}(3_{\bar b},4_{b},1_{\bar{q}},2_{q} ,5_{H}) \biggr) \,  \nonumber\\
& \hspace{4.5cm} -\delta_{i_2}^{\;\;\bar i_1} \delta_{i_4}^{\;\;\bar i_3}  \, \biggl( F^{(L)}(1_{\bar b},4_{b},3_{\bar{q}},2_{q} ,5_{H}) + F^{(L)}(3_{\bar b},2_{b},1_{\bar{q}},4_{q} ,5_{H}) \biggr) \bigg]\,.
\nonumber 
\end{align}

\subsection{Tree-Level Amplitudes}
\label{sec:trees}
The tree-level amplitudes can be obtained using the BCFW recursion relations~\cite{Britto:2004ap,Britto:2005fq} within the spinor helicity formalism. In the $\bar{b}bggH$ case, we choose to shift the momenta of gluons $3$ and $4$, while in the $\bar{b}b\bar{q}qH$ case we choose particles $1$ and $4$ to avoid shifting the momenta of adjacent quarks of the same flavour. Moreover, we ensure that the shifted brackets $[\hat{i}\ra, |\hat{j}]$ do not belong to particles of helicities $i^-, j^+$. These choices are necessary for the validity of the recursion relations as they prevent the shifted amplitude from having poles at infinity.

For the $\bar{b} b g g H$ channel we obtain the following non-vanishing tree-level partial amplitudes,
\begin{align} \label{eq:A0bbggh}
	A^{(0)}(1^+_{\bar{b}},2^+_{b},3^+_g,4^+_g,5_{H}) &=  \frac{s_5}{\spA 23 \spA 34 \spA 41} \,, \\
	A^{(0)}(1^+_{\bar{b}},2^+_{b},3^-_g,4^-_g,5_{H}) &= -\frac{\spB 12^2}{\spB 23\spB 34\spB 41} \,, \\
	A^{(0)}(1^+_{\bar{b}},2^+_{b},3^+_g,4^-_g,5_{H}) &= \frac{\spA 24 \spAB 451^2}{s_{234} \spA 23 \spA 34 \spAB 251}-\frac{s_5 \spB 13^3}{s_{134} \spB 14 \spB 34 \spAB 251} \,.
\end{align}
The $\bar{b},b$ quarks need to have the same helicity as that is the only way they can couple to the Higgs boson. For the $\bar{b} b g g H$ channel the ``all-plus'' and MHV configurations vanish, and we are left with
\begin{align} \label{eq:A0bbqqh}
	A^{(0)}(1^+_{\bar b},2^+_{b},3^+_{\bar q},4^-_{q},5_{H}) &=  \frac{\spAB 451^2}{s_{234} \spA 34 \spAB 251}+\frac{s_5 \spB 31^2}{s_{134} \spB 34 \spAB 251} \,. 
\end{align}

In both cases, due to the colour decomposition of the full amplitudes given by Eq.~\eqref{eq:colourdecomposition}, the $A^{(0)}(1^+,2^+,3^-,4^+,5_{H})$ partial amplitude is related to $A^{(0)}(1^+,2^+,3^+,4^-,5_{H})$ by swapping the particles $1\leftrightarrow 2,3\leftrightarrow 4$ , and flipping the overall sign for the subprocess $\bbggh$. The remaining non-vanishing helicity configurations can be obtained by parity transformations, that is by swapping the brackets $\la \phantom{1} \ra \leftrightarrow [\phantom{1}]$.

\section{Amplitude Reduction and Reconstruction}
\label{sec:reduction}

We derive the analytic form of the two-loop QCD helicity amplitudes using the framework discussed in Refs.~\cite{Badger:2021owl,Badger:2021imn}, which combines the Feynman diagram approach with numerical sampling over finite fields and functional reconstruction techniques. We generate a set of Feynman diagrams contributing to 
both the $\bbggh$ and $\bbqqh$ processes using 
\textsc{QGRAF}~\cite{Nogueira:1991ex}, and perform diagram filtering, topology identification and colour decomposition using a collection of in-house \textsc{Mathematica} and \textsc{Form}~\cite{Kuipers:2012rf,Ruijl:2017dtg} scripts. 
In the leading colour approximation there are 749 (264) Feynman diagrams contributing to the two-loop $\bbggh$ ($\bbqqh$) QCD amplitudes, including all closed fermion loop contributions.
The numerators of the loop amplitudes are then computed for each independent helicity configuration, and the spinor-helicity manipulations are carried out using the library \textsc{Spinney}~\cite{Cullen:2010jv}.
The helicity dependent loop numerators are written in terms of spinor products ($\spA{i}{j}$, $\spB{i}{j}$), 
scalar products ($p_i\cdot p_j$, $k_i\cdot k_j$, $k_i \cdot p_j$) and 
spinor strings ($\spAB{i}{k_i}{j}$, $\spAB{i}{p_5}{j}$, $\spAA{i}{k_i p_5}{j}$, $\spBB{i}{k_i p_5}{j}$), where $p_i$'s and $k_i$'s are the external and loop momenta, respectively. The angle $|i\rangle$ and square $|i]$ brackets denote the usual holomorphic and anti-holomorphic bi-spinors built from the external massless momenta. The momenta appearing between them in the spinor strings are intended as the matrices obtained by contracting the corresponding four-momenta with the vector of the Pauli matrices, e.g. $\spAB{i}{p_5}{j} = \spAB{i}{\sigma_{\mu}}{j} \, p^{\mu}_5$.
The loop numerators are therefore expressed as linear combinations of monomials of loop-momentum dependent objects 
that multiply coefficients which depend only on the external kinematics.
These coefficients contain spinor products and loop momentum independent spinor strings, 
and such objects not only are not independent  (due to momentum conservation and Schouten identity), 
but they are also incompatible with the finite field arithmetic framework.

In order to allow for the use of finite field sampling in the computation of the helicity amplitudes, 
we make use of an explicit rational parametrisation of the external kinematics. 
This is constructed with help of the momentum twistor~\cite{Hodges:2009hk} and spinor-helicity formalisms. To obtain a configuration for the off-shell five-particle configuration we begin with a massless configuration
for six particles. While the exact parametrisation is not important, the form presented in
Ref.~\cite{Badger:2016uuq} is a useful starting point. We can think of the massless process as the result of the
off-shell leg decaying into a pair of massless particles. There are 8 independent variables in the six-particle
process, but we can reduce the degrees of freedom by choosing one of the decay direction to be
collinear with one of the other massless legs in the five-point kinematics. Since the momentum
twistors are associated with complex momenta we consider the angle and square bracket spinors
products to be independent. We can write this explicitly as follows: the six massless momenta $q_i^\mu$ are related to the off-shell five particle momenta $p_i^\mu$ by
\begin{align}
  p_1^\mu = q_1^\mu\,, \ \qquad 
  p_2^\mu = q_2^\mu\,, \ \qquad 
  p_3^\mu = q_3^\mu\,, \ \qquad
  p_4^\mu = q_4^\mu \,, \ \qquad
  p_5^\mu = q_5^\mu+q_6^\mu \,.
\end{align}
We impose the constraints $\spA{q_2}{q_6} = 0 $ and $\spB{q_2}{q_6} = 0$ to reduce the independent variables to 6. 
We then apply a transformation onto the following choice of variables:
\begin{equation}
\begin{aligned}
  x_1 &= s_{12}, \\
  x_2 &= -\frac{\trp(1234)}{s_{12}s_{34}}, \\
  x_3 &= \frac{\trp(134152)}{s_{13}\trp{(1452)}}, \\
  x_4 &= \frac{s_{23}}{s_{12}}, \\
  x_5 &= -\frac{\trm(1(2+3)(1+5)523)}{s_{23}\trm(1523)}, \\
  x_6 &= \frac{s_{45}}{s_{12}},
\end{aligned}
\label{eq:mtvardefs}
\end{equation}
where $\tr_{\pm}(ij \cdots kl) = \frac{1}{2} \tr((1\pm\gamma_5)\slashed{p}_i\slashed{p}_j \cdots\slashed{p}_k\slashed{p}_l)$.
Some features of this parametrisation are particularly convenient. For example, the only dimensionful 
quantity is $x_1$, and $\tr_5$ is rational. However, such a choice does break some symmetries
in the problem and so different helicity configurations will have different complexities. For the
processes considered here the polynomial complexity was manageable using this form.
Applications to other off-shell five-particle processes may require further thought. 

Having set up a rational parametrisation of the external kinematics, the helicity dependent loop numerators are constructed analytically and
ready to be further processed.
This is the starting point of our numerical algorithm in the finite field setup. 
In order to write the loop amplitude in terms of Feynman scalar integrals, 
we first define the integral families for the \textit{maximal topologies}, 
which are the topologies with the maximum number of loop propagators (8 in this case). 
There are 15 maximal topologies for both the $\bbggh$ and $\bbqqh$ processes, and they are shown in Fig.~3 of Ref.~\cite{Badger:2021nhg}. 
Each diagram topology appearing in the amplitude can be mapped onto at least one of the maximal topologies. 
After that is done, the objects which depend on the loop momenta in the helicity dependent loop numerators are written in terms of 
the 11 propagators associated with the chosen maximal topology. 
These mapping procedures are performed numerically within the \textsc{FiniteFlow} framework~\cite{Peraro:2019svx}. 
External kinematics that involve in the mapping are already expressed in terms of momentum twistor variables.
At this stage, the loop amplitude is expressed as a linear combination of scalar integrals that will further be 
reduced to a set of master integrals. The coefficients associated with these scalar integrals are functions only of the external kinematics in the
form of momentum twistor variables.

The integrand is now ready to be reduced to a set 
of master integrals using the Integration-by-Parts (IBP) reduction method~\cite{Chetyrkin:1981qh} within the finite field setup. The IBP relations are generated using \textsc{LiteRed}~\cite{Lee:2012cn} and solved numerically over finite fields using the Laporta algorithm~\cite{Laporta:2001dd}. The IBP reduction is performed directly to a set of master integrals with uniform transcendental (UT) weight identified in Ref.~\cite{Abreu:2020jxa}. The UT master integrals are further decomposed into a basis of special functions $f$ that is built out of the UT master integral components as proposed in Ref.~\cite{Badger:2021nhg}. We refer to this basis of special functions as the \textit{master integral function basis}.
Subtracting the UV and IR poles from the bare helicity amplitudes and performing Laurent expansion in the dimensional regulator $\eps = (4-d)/2$, we arrive at the following expression for the finite remainders,
\begin{equation}
F^{(L)} = \sum_{i} r_i(x) m_i(f)\,,
\label{eq:finrem}
\end{equation}
where $m_i(f)$ are monomials of the special functions and $r_i$ are rational functions of the momentum twistor variables $x$. It is important to note that the definition of the UT master integrals involve three square roots. One is related to the pseudo-scalar invariant $\tr_5$, which captures the parity-odd part of the spinor expressions and is therefore present already in the rational coefficients. This square root is rationalised by the momentum-twistor inspired parameterisation~\eqref{eq:mtvardefs}. The other two square roots are associated with the three-mass triangle Gram determinants,
\begin{align}
  \Delta_1 &= 2 \sqrt{(p_{23} \cdot p_{5})^2 - p_5^2 s_{23}} \,,&
  \Delta_2 &= 2 \sqrt{(p_{12} \cdot p_{5})^2 - p_5^2 s_{12}} \,,
  \label{eq:squareroots}
\end{align}
where $p_{ij} = p_i + p_j$.
They are not rationalised by our parameterisation and may therefore be problematic for the finite field setup. We overcome this issue by absorbing the three square roots in the definition of the UT master integrals, which is possible because they appear only as overall normalisations of the latter. As a result, they are contained in the monomials $m_i(f)$ in Eq.~\eqref{eq:finrem} and do not appear in the amplitude reconstruction.
 We note that in our computation, while the $\tr_5$ originating from the UT master integral definition is absorbed in $m_i(f)$, 
the $\tr_5$ already present in the rational coefficients is rationalised and present in the amplitude reconstruction.

At this stage we have an algorithm which computes the coefficients of the special function monomials, $r_i(x)$ in Eq.~\eqref{eq:squareroots}, 
numerically over finite fields for each of the independent helicity configurations of the two processes $\bar{b}bggH$ and $\bar{b}b\bar{q}qH$. 
We emphasise that from the start of our numerical algorithm until the evaluation of $r_i(x)$, the computation is done within one system of \textsc{FiniteFlow} graphs.
The last step which remains to be done is the functional reconstruction of the rational coefficients. This task is made challenging by the high polynomial degrees, which are shown in the third column of Tables~\ref{tab:degrees2g2bH} and~\ref{tab:degrees2q2bH}. We tackle the complexity of the reconstruction through the strategy already used in Refs.~\cite{Badger:2021nhg,Badger:2021imn}. We refer to Ref.~\cite{Badger:2021imn} for a thorough discussion, and give here only a brief outline. 

The first step of the strategy consists in fitting the linear relations among the rational coefficients. These linear relations are then used to express the rational coefficients in terms of a set of linearly independent coefficients, which are chosen to have the lowest possible degrees. The degrees of the independent rational coefficients are given in the fourth column of Tables~\ref{tab:degrees2g2bH} and~\ref{tab:degrees2q2bH}.

The second step of our reconstruction strategy consists in determining the factors in the denominators of the rational coefficients. The analytic structure of the special functions is determined by a set of algebraic functions of the kinematics called letters. In other words, the singularities and branch cuts of the special functions are located on the hypersurfaces where any of the letters vanishes (or goes to infinity). It is therefore natural to expect that the rational coefficients which multiply the special functions should feature poles which are similarly linked to the letters, given for this problem in Ref.~\cite{Abreu:2020jxa}. Inspired from the letters we can therefore guess the factors in the denominators of the rational coefficients. Our guess is given by
\begin{equation}
\begin{aligned}
\biggl\{  & \spA{i}{j} \,,  \spB{i}{j}  \,, \spAB{i}{p_5}{j}  \,, s_{ij} \,, s_{ij} - s_{kl} \,, 
  s_{i5} - p_5^2 \,, p_5^2 \,, \tr_5 \, , \Delta_1 \,, \Delta_2 \, , \\
&  s_{15} (s_{13}+s_{34}) - p_5^2  s_{34}  \,,
 s_{25} (s_{24}+s_{34}) - p_5^2  s_{34} \biggr\} \,,
\end{aligned}
\label{eq:coeffansatz}
\end{equation}
where the indices $i,j$ in the spinor expressions vary from $1$ to $4$, while the indices $i,j,k,l$ in the Mandelstam invariants vary from $1$ to $5$. The various free indices in each of the factors are understood to be different from each other. We then match an ansatz made of the factors in Eq.~\eqref{eq:coeffansatz} against the rational coefficients reconstructed on a random univariate slice of the phase space. This allows to determine entirely the denominators, as well as some factors in the numerators.

Having determined the denominators of the rational coefficients we can proceed to the third and last step of our reconstruction strategy:  a univariate partial fraction decomposition on the fly. We find that, with the parameterisation of the external kinematics given by Eq.~\eqref{eq:mtvardefs}, partial fractioning with respect to $x_5$ is the most convenient choice. 
The partial fraction decomposition of multivariate functions has recently drawn a lot interest as a powerful tool to simplify the analytic expressions of scattering amplitudes. A number of new approaches have been proposed, which make use of algebraic geometry techniques to handle the multivariate case efficiently~\cite{Leinartas:1978,Raichev:2012,Abreu:2019odu,Boehm:2020ijp,Heller:2021qkz}. These algorithms are however applied to known analytic expressions, namely after the rational reconstruction has been performed. Conversely, the univariate partial fraction decomposition can be performed within the finite field setup to simplify the reconstruction of the rational coefficients.
The algorithm makes use of the knowledge of the denominators to construct an ansatz for the decomposition into partial fractions with respect to the chosen variable. The coefficients of this ansatz can then be solved for and reconstructed within the finite field workflow. In order to reconstruct the coefficients of the partial fraction decomposition we perform a further matching of the factors from the ansatz~\eqref{eq:coeffansatz}. The determined factors are then removed, and the remaining functions are reconstructed with \textsc{FiniteFlow}'s reconstruction algorithm. The degrees of the rational coefficients which remain to be reconstructed are shown in the fifth column of Tables~\ref{tab:degrees2g2bH} and~\ref{tab:degrees2q2bH}. Note that after partial fractioning the denominators are not entirely determined from the ansatz~\eqref{eq:coeffansatz}. The univariate partial fraction decomposition in fact introduces spurious factors in the denominators. The latter could be determined as well, but we refrain from doing so as it does not reduce the complexity of the reconstruction.

\renewcommand{\arraystretch}{1.5}
\begin{table}[t!]
\centering
\begin{tabularx}{0.95\textwidth}{|C{0.8}|C{1.2}|C{0.8}|C{1.0}|C{1.2}|C{1.0}|}
\hline
 $\bbggh$     & helicity configurations & $r_i(x)$ & independent $r_i(x)$ & partial fraction in $x_5$ & number of points \\
\hline
$F^{(2),1}$ & $++++$ & 63/57   & 52/46   & 20/6  & 3361 \\
            & $+++-$ & 135/134 & 119/120 & 28/22 & 24901 \\
            & $++--$ & 105/111 & 105/111 & 22/12 & 4797 \\
\hline
$F^{(2),n_f}$ & $++++$ & 45/41 & 45/41 & 16/6 & 1381 \\
              & $+++-$ & 94/95 & 94/95 & 17/6 & 1853 \\
              & $++--$ & 89/95 & 62/69 & 18/3 & 2492 \\
\hline
$F^{(2),n_f^2}$ & $++++$ & 12/8  & 9/7   & 0/0 & 3 \\
                & $+++-$ & 11/16 & 11/16 & 3/0 & 22 \\
                & $++--$ & 12/20 & 8/16  & 8/0 & 242 \\
\hline
\end{tabularx}
\caption{\label{tab:degrees2g2bH} Maximum numerator/denominator polynomial degrees of the finite remainder coefficients $r_i(x)$ in Eq.~\eqref{eq:finrem} 
at each stage of our reconstruction steps, together with the number of sample points needed for the analytic reconstruction in the $\bbggh$ subprocess, for the various closed fermion loop contributions.}
\end{table}

\begin{table}[t!]
\centering
\begin{tabularx}{0.95\textwidth}{|C{0.8}|C{1.2}|C{0.8}|C{1.0}|C{1.2}|C{1.0}|}
\hline
 $\bbqqh$     & helicity configurations & $r_i(x)$ & independent $r_i(x)$ & partial fraction in $x_5$ & number of points \\
\hline
$F^{(2),1}$ & $+++-$ & 82/81 & 69/70 & 24/16 & 10326 \\
\hline
$F^{(2),n_f}$ & $+++-$ & 28/30 & 25/24 & 8/6 & 379 \\
\hline
$F^{(2),n_f^2}$ & $+++-$ & 6/11  & 6/11  & 3/0 & 22 \\
\hline
\end{tabularx}
\caption{\label{tab:degrees2q2bH} Maximum numerator/denominator polynomial degrees of the finite remainder coefficients $r_i(x)$ in Eq.~\eqref{eq:finrem} 
at each stage of our reconstruction steps, together with the number of sample points needed for the analytic reconstruction in the $\bbqqh$ subprocess, for the various closed fermion loop contributions.}
\end{table}

Following the strategy outlined above we reconstructed the partial finite remainders for the independent mostly-plus helicity configurations of the subprocesses $\bbggh$ and $\bbqqh$. Reconstruction data are shown in Tables~\ref{tab:degrees2g2bH}~and~\ref{tab:degrees2q2bH}. The mostly-minus helicity finite remainders can be obtained by parity conjugation. Moreover, the finite remainders for the helicity configuration $++-+$ can be obtained by swapping $1\leftrightarrow2$, $3\leftrightarrow 4$ in the $+++-$ finite remainders, as discussed in Section~\ref{sec:trees} for the tree-level amplitudes. 
This symmetry follows from the colour structure and thus holds at any loop order. 
We could therefore evaluate the $++-+$ finite remainders by permuting numerically the $+++-$ ones, as we do in order to get all the other helicity configurations from the independent ones. By permuting numerically we mean that we obtain the values of the permuted finite remainders by evaluating the un-permuted ones at permuted points. This is possible because our approach to the evaluation of the special functions, which we discuss in Section~\ref{sec:Hbasis}, handles the analytic continuation to any region automatically. Each numerical permutation therefore amounts to one more evaluation for each point. The permutation that takes from $+++-$ to $++-+$ is however peculiar, as it is covered by the basis of special functions defined in Ref.~\cite{Badger:2021nhg}. We can thus reconstruct the finite remainders for both helicity configurations directly in terms of the same basis of special functions. This is much more convenient, as it reduces the number of permutations which need to be carried out numerically, this way decreasing the global evaluation time of the finite remainders. For this reason we reconstructed the analytic expression of the $++-+$ finite remainders as well, and include the $++-+$ configuration in the independent helicity set in the following sections. 
The relation with the $+++-$ configuration constitutes a non-trivial check of our results, which we discuss in Section~\ref{sec:validation}.

\section{A Custom Basis of Special Functions for the Finite Remainders}
\label{sec:Hbasis}

The one and two-loop finite remainders are expressed as combinations of rational coefficients --~functions of the momentum twistor variables~\eqref{eq:mtvardefs}~-- and monomials of square roots and elements of the master integral function basis $\{f^{(w)}_i\}$. The latter were classified in Ref.~\cite{Badger:2021nhg} so as to span the cyclic permutations of the planar five-particle integrals with one massive off-shell leg up to two loops. The function space of the finite remainders is however simpler than that of the integrals and of the amplitudes. This becomes particularly clear when we express the special functions in terms of Chen's iterated integrals~\cite{Chen:1977oja} (see the notes~\cite{Brown:2013qva} for a thorough discussion of their properties). Given a set of logarithmic integration kernels $\left\{ d\log W_i \right\}$, where the arguments $W_i$ are algebraic functions of the external kinematics, Chen's iterated integrals can be defined iteratively as
\begin{align} \label{eq:Chen}
\left[W_{i_1}, \ldots, W_{i_n}  \right]_{s_0} (s) = \int_0^1 dt \, \frac{d \log W_{i_n}\left(\gamma(t)\right)}{dt} \left[W_{i_1}, \ldots, W_{i_{n-1}}  \right]_{s_0} \left(\gamma(t)\right) \,,
\end{align}
where we denote cumulatively by $s$ the dependence on the external kinematics, $s_0$ is an arbitrary reference point, and $\gamma$ is an arbitrary contour in the phase space going from $\gamma(0)=s_0$ to $\gamma(1)=s$. The iteration starts with $[]_{s_0}(s) := 1$, and the number of iterated integrations --~$n$ in Eq.~\eqref{eq:Chen}~-- is called transcendental weight. At two loops up to order $\eps^0$ iterated integrals with transcendental weight up to four are required. In our case, the integration kernels are given by the letters $\{W_i\}$ of the alphabet defined in Ref.~\cite{Abreu:2020jxa}. 
Although they may look rather abstract, Chen's iterated integrals offer several important advantages. Most importantly, they implement automatically all the functional relations which would otherwise be hidden in a representation in terms of other types of functions, such as Goncharov polylogarithms. This property was exploited in Ref.~\cite{Badger:2021nhg} to construct the master integral function basis $\{f^{(w)}_i\}$. 
The expression in terms of Chen's iterated integrals of the master integral basis functions can be obtained through their definition in terms of master integrals components given in Ref.~\cite{Badger:2021nhg}, and the canonical differential equations for the master integrals given in Ref.~\cite{Abreu:2020jxa} (see Ref.~\cite{Badger:2021nhg} for a thorough discussion). Another important benefit of using iterated integrals is that their analytic structure is beautifully manifest: an iterated integral may have singularities or branch points only where one of its letters vanishes or diverges. Expressing the special functions in the finite remainders in terms of Chen's iterated integrals therefore highlights their analytic structure. Indeed this unveils important simplifications: certain letters, which are present in the expressions of the master integrals, are absent in the finite remainders. In other words, certain branch cuts of the integrals drop out of the finite remainders. We observe the same pattern noticed for the $Wb\bar{b}$ amplitudes in Ref.~\cite{Badger:2021nhg}. One letter, $W_{49} = \text{tr}_5$, is present in the bare amplitudes but absent in the finite remainders. This pattern of the pseudo-scalar invariant $\text{tr}_5$ has already been observed explicitly in many massless two-loop five-particle finite remainders~\cite{Badger:2018enw,Abreu:2018zmy,Abreu:2018aqd,Chicherin:2018yne,Chicherin:2019xeg,Abreu:2019rpt,Abreu:2019odu,Badger:2019djh,Caron-Huot:2020vlo,Abreu:2020cwb,Chawdhry:2020for}, and is linked to an underlying cluster algebra structure~\cite{Chicherin:2020umh}.
In addition, six letters, $W_i$ with $i=16,17,27,28,29,30$, are present in the two-loop integrals but drop out already at the level of the bare amplitudes (truncated at order $\eps^0$ at two loops). It would be of great interest to uncover the physical principle underlying these simplifications.

The simplifications of the analytic structure mentioned above require the interplay among various elements of the master integral function basis $\{f^{(w)}_i\}$ of Ref.~\cite{Badger:2021nhg}. In other words, the separate terms of the finite remainders may have spurious branch cuts which cancel out in the sum. It is therefore convenient to construct a new, ad hoc basis of special functions where the properties of the finite remainders are manifest. In addition to being more elegant from the theoretical point of view, such a basis is also much more convenient from the practical point of view. Evaluating the master integral function basis $\{f^{(w)}_i\}$ in fact requires handling integration kernels --~letters~-- which eventually cancel out in the objects we are interested in evaluating. It is desirable that these cancellations take place analytically rather than numerically, so that the spurious kernels are avoided altogether. For this reason we need a new basis of special functions where these properties are manifest, and an approach to evaluate it which by-passes the master integral function basis $\{f^{(w)}_i\}$ and its unnecessary complexity.

We thus construct a new basis of special functions, which we label by $\{h^{(w)}_i\}$, where $w=1,2,3,4$ denotes the transcendental weight. This new basis contains only those linearly independent combinations of functions $f^{(w)}_i$ which are actually present in the finite remainders, and whose iterated integral expression is free of the letters $W_i$ with $i=16,17,27,28,29,30,49$. We dub this basis \textit{finite remainder function basis}, as opposed to the master integral function basis $\{f^{(w)}_i\}$. This leads to an important simplification in the expressions of the finite remainders, which are then expressed as combinations of rational coefficients and monomials in the $h^{(w)}_i$'s and the square roots. In particular, only $23$ weight-4 functions $\{h^{(4)}_i\}_{i=1}^{23}$ are required, to be compared with the $113$ weight-4 functions in the master integral function basis. Since the evaluation of the weight-4 functions is the most expensive step in the evaluation of the finite remainders, this reduction has a strong positive impact on the total evaluation time. Note that we have also improved the master integral basis $\{f^{(w)}_i\}$ of Ref.~\cite{Badger:2021nhg} by identifying relations among the higher weight functions and products of lower weight ones which were originally missed. We achieved this following the approach of Ref.~\cite{Badger:2021nhg}, but evaluating the boundary values with higher accuracy.

In order to evaluate numerically the finite remainder function basis $\{h^{(w)}_i\}$ we apply the method of the generalised power series expansion~\cite{Francesco:2019yqt}. This approach has already found several successful applications to the evaluation of master integrals from the differential equations they satisfy~\cite{Francesco:2019yqt,Bonciani:2019jyb,Frellesvig:2019byn,Abreu:2020jxa,Becchetti:2020wof,Bonciani:2021zzf,abreu2021twoloop}. Here, following Ref.~\cite{Badger:2021nhg}, we apply it directly to the basis of special functions. The main advantage is that having a basis of special functions allows to subtract from the bare amplitudes the IR and UV poles analytically, which makes it possible to reconstruct the finite remainders directly over finite fields. 

The method of the generalised power series expansion can be applied to the finite remainder and master integral function basis because they too, like the master integral bases they stem from, satisfy systems of differential equations in the canonical form~\cite{Henn:2013pwa}. This follows from the fact that the functions in the bases $\{h^{(w)}_i\}$ and $\{f^{(w)}_i\}$ are by construction \textit{pure}, that is they have the following two properties. First, they have uniform transcendental weight, i.e. they are expressed by linear combinations of Chen's iterated integrals having the same transcendental weight with constant rational coefficients. Explicitly, any of the functions $h^{(w)}_i$, for $w>0$, is given in terms of Chen's iterated integrals by
\begin{align} \label{eq:example_h}
h^{(w)}_{i}(s) = \sum_{I = (i_1,\ldots,i_w)} c_I^{(i)} \, \left[W_{i_1}, \ldots, W_{i_w}  \right]_{s_0} (s) \, ,
\end{align}
for some rational constant coefficients $c_I^{(i)}$\footnote{In general the iterated integrals in Eq.~\eqref{eq:example_h} include also transcendental constants, such as $\pi$ and $\zeta(3)$. We neglect them here to simplify the presentation.}.
Second, the derivative of a pure weight-$w$ function is a uniform transcendental function with weight $w-1$. This follows straightforwardly from the differential of Chen's iterated integrals,
\begin{align} \label{eq:ChenDerivative}
d \left[W_{i_1}, \ldots, W_{i_n}  \right]_{s_0} (s) = d \log W_{i_n}(s) \, \left[W_{i_1}, \ldots, W_{i_{n-1}}  \right]_{s_0} (s) \,,
\end{align}
where we note that the derivatives of $\log W_i$ are algebraic functions and hence have transcendental weight $0$.
The transcendental purity of the functions in the basis therefore implies that the vector of all weight-$w$ functions in the finite remainder basis, $\vec{h}^{(w)}= \bigl(h^{(w)}_1, h^{(w)}_2, \ldots \bigr)^T$, satisfies a differential equation of the form
\begin{align} \label{eq:dhw}
d \vec{h}^{(w)} = \left( \sum_{j} b_j^{(w)} \, d\log W_j  \right) \,  \vec{h}^{(w-1)} \,,
\end{align}
for $w>1$, where $b_j$ are constant rational matrices. The matrices $b_j^{(w)}$ can be determined easily from the iterated integral expression of the functions through Eq.~\eqref{eq:ChenDerivative}. 
In general the derivatives of the weight-$w$ functions may involve weight-$(w-1)$ functions which are not needed to express the finite remainders. These must be included in the basis as well, in order to be able to write down the differential equations. So, strictly speaking, the finite remainder function basis $\{h^{(w)}_i\}$ contains all linearly independent functions which appear in the finite remainders and in the derivatives of the higher-weight functions of the basis itself.
The differential equations~\eqref{eq:dhw} for all weights can be put together in one system by defining the vector of all the functions in the finite remainder basis as
\begin{align}
\vec{h} = \left( \eps^4 \vec{h}^{(4)}\,, \eps^3 \vec{h}^{(3)}\,, \eps^2 \vec{h}^{(2)} \,, \eps \vec{h}^{(1)} \,, 1 \right)^T \,,
\end{align}
where $\eps$ is an auxiliary parameter with transcendental weight $-1$, so that the vector $\vec{h}$ is pure with transcendental weight $0$. The full basis $\vec{h}$ then satisfies a system of differential equations in the canonical form~\cite{Henn:2013pwa},
\begin{align} \label{eq:dh}
d \vec{h} = \eps \, \left(  \sum_{i=1}^{49} a_i \, d\log W_i \right) \, \vec{h} \,,
\end{align}
where the constant matrices $a_i$ are strictly upper block triangular, with the blocks given by the matrices $b_i^{(w)}$ in the differential equations~\eqref{eq:dhw} for the various weights.
This system of differential equations is however much simpler than that for the master integrals. First, it contains only the letters which do appear in the finite remainders, i.e. $a_i = 0$ for $i=16,17,27,28,29,30,49$. Second, while the differential equations for the master integrals contain information about all the orders in $\eps$, the system~\eqref{eq:dh} encodes only those orders which are relevant for the finite remainders, i.e. up to $\eps^4$. The constant matrices $a_i$ are in fact nilpotent with degree $5$, i.e. $(a_i)^5 = 0$ for any $i$, which follows from their strictly upper triangular form.

We evaluate the finite remainder function basis $\{h^{(w)}_i\}$ by solving the system of differential equations~\eqref{eq:dh} numerically with the method of the generalised power series expansions~\cite{Francesco:2019yqt}. For this purpose we make use of the public \textsc{Mathematica} package \textsc{DiffExp}~\cite{Hidding:2020ytt}. In order to compute the boundary values, i.e. the values of the functions at some base point, we relate the finite remainder functions to master integral components using the definition of the master integral function basis~\cite{Badger:2021nhg}, and evaluate the latter through their analytic expression in terms of Goncharov polylogarithms~\cite{2011arXiv1105.2076G,Remiddi:1999ew,2001math......3059G} provided in Refs.~\cite{Papadopoulos:2015jft,Syrrakos:2020kba,Canko:2020ylt}. We evaluate the Goncharov polylogarithms numerically with the \textsc{C++} library \textsc{GiNaC}~\cite{Vollinga:2004sn}. We provide in ancillary files the differential equations~\eqref{eq:dh} and the values of the finite remainder functions at six points, one for each $2\to 3$ scattering region with massless incoming particles, with (at least) $200$-digit accuracy. Using this information, the generalised power series expansion method allows to evaluate the finite remainder function basis reliably anywhere in the phase space. This technique in fact handles the analytic continuation automatically, so that also the permutations required to evaluate the complete finite remainders starting from the partial ones --~as shown in Eqs.~\eqref{eq:colourdecomposition} for the amplitudes~-- as well as the other helicity configurations can be performed numerically straightforwardly. We assessed the reliability of this evaluation approach by checking the convergence of the finite remainders close to the spurious poles of the rational coefficients. We discuss this and other checks in the next section.

\section{Further Validation}
\label{sec:validation}

The finite remainders are defined by subtracting the expected UV and IR poles from the bare amplitudes. Therefore, the fact that all the poles in $\eps$ cancel out, so that our expressions for the finite remainders are indeed finite, is already a strong check of the validity of our results. We have also checked that our amplitudes are correctly normalised by comparing the spin and colour averaged squared tree-level amplitudes with full colour dependence against \textsc{MadGraph}~\cite{Alwall:2014hca}.
On top of that, we performed a number of additional checks, which we discuss in the next subsections.

\subsection{Direct computation of the squared finite remainders} 

We checked the helicity amplitudes derived in this paper against a squared matrix element computation, carried out independently following the approach used in the previous work on $u\bar{d}\to Wb\bar{b}$~\cite{Badger:2021nhg}.
In the squared matrix element computation the two-loop amplitude is interfered with the tree-level amplitude analytically. After manipulating the
Dirac traces, the loop numerators contain only scalar products ($k_i \cdot k_j$, $k_i \cdot p_j$, $p_i \cdot p_j$), that can be mapped easily onto propagator denominators. This way we derive an analytic form of the two-loop squared matrix element written as a linear combination of scalar Feynman integrals. The squared matrix element is then processed further through IBP reduction to obtain the master integral representation, followed by mapping of the master integrals onto the basis of special functions, subtraction of UV and IR singularities, and finally Laurent expansion in $\eps$. 
All these steps are carried out numerically over finite fields within the \textsc{FiniteFlow} framework. 
The squared matrix element computation uses the Conventional Dimensional Regularisation (CDR) scheme, where internal and external live in $d=4-2 \eps$ dimensions. We find complete agreement with this approach and the helicity amplitudes in the tHV scheme at the level of the squared finite remainders.

\subsection{Renormalisation scale dependence}

The analytic expressions of the one- and two-loop finite remainders have been derived with the renormalisation scale $\mu$ set to $\mu  = 1$. 
The dependence of the finite remainders on the renormalisation scale can be determined by restoring the $\mu$ dependence of the strong and Yukawa couplings ($\alpha_s \rightarrow \alpha_s \mu^{2\eps}$, 
$y_b \rightarrow y_b \mu^{\eps}$), and by replacing $(-s_{ij})^{-\eps} \rightarrow (-\mu^2/s_{ij})^\eps$ in the $I_1$ operators defined in Eqs.~\eqref{eq:I1bbqqh}~--~\eqref{eq:I1bbggh}, which enter the pole terms in Eqs.~\eqref{eq:poles1L}~--~\eqref{eq:poles2L}. In order to capture the scale dependence of the finite remainders we define the difference
\begin{equation}
\delta F^{(L),i}(\mu^2) = F^{(L),i}(\mu^2) - F^{(L),i}(\mu^2 = 1) \,,
\end{equation}
where the dependence on the kinematic variables is understood.
The difference $\delta F^{(L),i}(\mu^2)$ is entirely determined by the finite remainders evaluated at $\mu^2=1$ --~which we reconstructed analytically~-- and logarithms of $\mu^2$.
For the $\bbggh$ finite remainders it is given by
\begingroup
\allowdisplaybreaks
\begin{align}
\delta F^{(1),1}(\mu^2)   & = \frac{31}{6} A^{(0)} \log(\mu^2) \,, \\
\delta F^{(1),n_f}(\mu^2) & = - \frac{2}{3} A^{(0)} \log(\mu^2) \,, \\
\delta F^{(2),1}(\mu^2) & = \log(\mu_R^2) \bigg\lbrace \bigg(\frac{1261}{54} - \frac{11}{72}\pi^2 + 9 \zeta_3\bigg) A^{(0)} + \frac{53}{6}F^{(1),1}(1) \bigg\rbrace
                             + \frac{1643}{72} A^{(0)} \log^2(\mu^2) \,, \\
\delta F^{(2),n_f}(\mu^2) & = \log(\mu_R^2) \bigg\lbrace \bigg(-\frac{241}{27} + \frac{\pi^2}{36} \bigg) A^{(0)} 
                                                                 - \frac{4}{3}F^{(1),1}(1)  + \frac{53}{6}F^{(1),n_f}(1) \bigg\rbrace \nonumber \\
                            & \quad - \frac{115}{18} A^{(0)} \log^2(\mu^2)  \,, \\
\delta F^{(2),n_f^2}(\mu^2) & = \log(\mu^2)\bigg(\frac{20}{27} A^{(0)} - \frac{4}{3} F^{(1),n_f}(1)\bigg) + \frac{4}{9} A^{(0)} \log^2(\mu^2) \,, 
\end{align}
\endgroup
while for the $\bbqqh$ finite remainders it is given by
\begingroup
\allowdisplaybreaks
\begin{align}
\delta F^{(1),1}(\mu^2)   & =  \frac{31}{6} A^{(0)} \log(\mu^2) \,, \\
\delta F^{(1),n_f}(\mu^2) & =  - \frac{2}{3} A^{(0)} \log(\mu^2) \,, \\
\delta F^{(2),1}(\mu^2) & = \log(\mu^2) \bigg\lbrace \bigg(\frac{5093}{216} - \frac{11}{12}\pi^2 + 14 \zeta_3\bigg) A^{(0)} 
                              + \frac{53}{6}F^{(1),1}(1) \bigg\rbrace  + \frac{1643}{72} A^{(0)} \log^2(\mu^2)  \,, \\
\delta F^{(2),n_f}(\mu^2) & = \log(\mu^2) \bigg\lbrace \bigg(-\frac{329}{54} + \frac{\pi^2}{6} \bigg) A^{(0)} 
                               - \frac{4}{3}F^{(1),1}(1)  + \frac{53}{6}F^{(1),n_f}(1) \bigg\rbrace \nonumber  \\
                            & \quad - \frac{115}{18} A^{(0)} \log^2(\mu^2) \,, \\
\delta F^{(2),n_f^2}(\mu^2) & = - \frac{4}{3} F^{(1),n_f}(1) \log(\mu^2) + \frac{4}{9} A^{(0)} \log^2(\mu^2) \, . 
\end{align}
\endgroup

To check that our results for the finite remainders have the correct scale dependence, we evaluate them at two kinematic points related by a rescaling by some positive factor $a$,
\begin{equation}
\begin{aligned}
\vec{s} &= (s_{12},s_{23},s_{34},s_{45},s_{15},p_5^2) \,,  \\ 
\vec{s}^{\,'} &= a \, \vec{s} = (a s_{12},a s_{23},a s_{34},a s_{45},a s_{15},a p_5^2) \,. 
\end{aligned}
\end{equation}
We then verify numerically that the finite remainders satisfy the following relation,
\begin{equation}
  \frac{F^{(L),i}(1,a \, \vec{s})+\delta F^{(L),i}(a,a \, \vec{s})}{A^{(0)}(a\,\vec{s})} 
= \frac{F^{(L),i}(1,\vec{s})}{A^{(0)}(\vec{s})} \,,
\end{equation}
where we extended the notation of the finite remainders and tree-level amplitudes to take into account their dependence on the phase space point $\vec{s}$.

\subsection{Relation between $+++-$ and $++-+$} 

The partial finite remainders for the single-minus helicity configurations, $+++-$ and $++-+$, are related by a permutation of the external particles,
\begin{equation} \label{eq:relations}
\begin{aligned} 
 & F^{(L)}(1^+_{\bar{b}},2^+_{b},3^-_g,4^+_g,5_{H}) = - \, F^{(L)}(2^+_{\bar{b}},1^+_{b},4^+_g,3^-_g,5_{H}) \,, \\
 & F^{(L)}(1^+_{\bar{b}},2^+_{b},3^-_{\bar{q}},4^+_q,5_{H}) = F^{(L)}(2^+_{\bar{b}},1^+_{b},4^+_{\bar{q}},3^-_q,5_{H}) \,.
\end{aligned}
\end{equation}
Permuting the special functions is however non-trivial, and may in general require analytic continuation. The permuted special functions are defined by
\begin{align}
\left(\sigma \circ \vec{h} \right) (\{p_i\}) = \vec{h} ( \{p_{\sigma(i)}\} )\, ,
\end{align}
where $\sigma$ denotes the permutation $(12345)\to(21435)$ of the external particles, and $\{p_i\}$ denotes the dependence on the external momenta. In order to check the relations~\eqref{eq:relations} analytically, we need to express the permuted functions $(\sigma \circ \vec{h} )$ in terms of the ones in the standard orientation, $\vec{h}$. To perform this operation in an algorithmic and robust way we resort to the system of differential equations~\eqref{eq:dh} satisfied by the finite remainder function basis. The permuted functions in fact satisfy the permuted differential equations
\begin{align}
d ( \sigma \circ \vec{h} ) = \eps \left[ \sum_{i=1}^{49} a_i \, d \log\left(\sigma \circ W_i \right) \right] \,  ( \sigma \circ \vec{h} ) \,.
\end{align}
Permuting the letters $W_i$ is however trivial, as they only involve algebraic functions. Since the alphabet is closed by construction under this permutation, we obtain an explicit system of differential equations for the permuted special functions, 
\begin{align} \label{eq:dsigmah}
d ( \sigma \circ \vec{h} ) = \eps \left( \sum_{i=1}^{49} a_i' \, d \log W_i  \right) \,  ( \sigma \circ \vec{h} ) \,.
\end{align}
The latter can straightforwardly be solved in terms of Chen's iterated integrals (see Ref.~\cite{Badger:2021nhg} for a thorough discussion). In order to be comparable with the solution of the system~\eqref{eq:dh}, which defines the finite remainder function basis $\vec{h}$, we must make sure that the same boundary point is used when solving both systems of differential equations in terms of iterated integrals. The boundary values can be obtained numerically with arbitrary precision using the expressions in terms of Goncharov polylogarithms of Refs.~\cite{Papadopoulos:2015jft,Syrrakos:2020kba,Canko:2020ylt}, as discussed at the end of Section~\ref{sec:Hbasis}. Using the differential equations~\eqref{eq:dsigmah} we can then express the permuted finite remainder special functions in terms of the ones in the original orientation through Chen's iterated integrals. Once that is done, the right-hand sides of Eqs.~\eqref{eq:relations} are written in terms of the same special function basis as the left-hand sides. Since the rational coefficients can be permuted analytically trivially, it is immediate to verify that our results for the one- and two-loop finite remainders satisfy the relations given by Eqs.~\eqref{eq:relations}.

\subsection{Convergence near spurious poles}

The rational coefficients in the finite remainders contain spurious poles, namely poles which are not related to the physical singularities of the amplitudes. When evaluating at a phase space point infinitesimally close to a spurious pole --~but at a finite distance from all physical poles~-- the rational coefficients diverge, whereas the finite remainders must stay finite. This can only occur through large numerical cancellations involving both the rational coefficients and the special functions. Verifying this behaviour constitutes both a check on our analytic expressions for the finite remainders and a stress test of our evaluation approach for the special functions. We do it as follows.
First, we use the factor matching strategy discussed in Section~\ref{sec:reduction} with the ansatz given by Eq.~\eqref{eq:coeffansatz}. This allows us to determine that the spurious poles in the rational coefficients are associated with the following factors in their denominators,
\begin{equation} \label{eq:spurious}
\begin{aligned}
\left\{P_i \right\} = \biggl\{ & \langle i | p_{5}| j] \,, \text{tr}_5 \,, \Delta_1 \,, \Delta_2 \,, s_{15} - s_{23} \,, s_{15} - s_{24} \,, s_{13} - s_{25} \,, s_{14} - s_{25} \,, s_{15} - s_{34} \,,  \\
  &  s_{25} - s_{34} \,, s_{14} - s_{35} \,, s_{23} - s_{45} \,,
 s_{15} (s_{13}+s_{34}) - p_5^2  s_{34}  \,,
 s_{25} (s_{24}+s_{34}) - p_5^2  s_{34} \biggr\} \,,
\end{aligned}
\end{equation}
where $i,j \in \{1,2,3,4\}$ with $i\neq j$. Next, for each factor in Eq.~\eqref{eq:spurious}, we construct a one-dimensional slice of the phase space, parametrised by a parameter $\delta$, such that $P_i = \delta$. As we probe the small-$\delta$ region all the other factors in Eq.~\eqref{eq:spurious} and the factors associated with the physical singularities --~$(p_i+p_j)^2$ and $p_i \cdot p_j$~-- must stay finite, i.e. they must neither diverge nor vanish. This ensures that we target a specific spurious pole, rather than multiple at once, and that we stay away from the physical singularities. Finally, we evaluate the finite remainders on these slices for increasingly small values of $\delta$. We evaluate the special functions with $64$-digit accuracy. Our analytic expressions for the finite remainders exhibit the expected behaviour: as $\delta$ approaches zero the rational coefficients diverge, while the finite remainders converge to finite values.

\section{Results}
\label{sec:results}

We provide the analytic expressions of the independent one- and two-loop finite remainders in the ancillary files. We present them as combinations of linearly independent rational coefficients and monomials of the square roots and of the finite remainder basis functions. The rational coefficients are expressed in terms of the momentum twistor variables defined in Eq.~\eqref{eq:mtvardefs}. We evaluate the finite remainder function basis numerically by integrating the differential equations~\eqref{eq:dh} with the method of the generalised series expansions. We also provide \textsc{Mathematica} scripts which illustrate how to evaluate the finite remainders interfered with the tree-level amplitudes for all the partonic channels contributing to the process $pp\to b\bar{b} H$, which we label as
\begin{equation}
\label{eq:channel_definition}
\begin{aligned}
&\mathbf{gg}:       \quad  g(-p_3) + g(-p_4) \rightarrow \bar{b}(p_1) + b(p_2) + H(p_5) \,, \\
&\mathbf{q\bar{q}}: \quad  q(-p_3) + \bar{q}(-p_4) \rightarrow \bar{b}(p_1) + b(p_2) + H(p_5) \,, \\
&\mathbf{\bar{q}q}: \quad \bar{q}(-p_3) + q(-p_4) \rightarrow \bar{b}(p_1) + b(p_2) + H(p_5) \,, \\
&\mathbf{b\bar{b}}: \quad b(-p_3) + \bar{b}(-p_4) \rightarrow \bar{b}(p_1) + b(p_2) + H(p_5) \,, \\
&\mathbf{\bar{b}b}: \quad \bar{b}(-p_3) + b(-p_4) \rightarrow \bar{b}(p_1) + b(p_2) + H(p_5) \,, \\
&\mathbf{bb}: \quad b(-p_3) + b(-p_4) \rightarrow b(p_1) + b(p_2) + H(p_5) \,, \\
&\mathbf{\bar{b}\bar{b}}: \quad \bar{b}(-p_3) + \bar{b}(-p_4) \rightarrow \bar{b}(p_1) + \bar{b}(p_2) + H(p_5) \,.
\end{aligned}
\end{equation}
The interference between the finite remainders and the tree-level amplitudes summed over colour and helicity is given at leading colour by 
\begin{align}
\sum_{\text{colour}} \sum_{\text{helicity}} \mathcal{A}^{(0) *} \mathcal{F}^{(L)} =: g_s^4 y_b^2 n^L N_c^{\alpha} \, \mathcal{H}^{(L)}\,,
\end{align}
where $\alpha=3$ for $\mathbf{gg}$ and $\alpha=2$ for all the other channels, and the \textit{reduced squared finite remainders}, $\mathcal{H}^{(L)}$, are given explicitly for each channel by
\begingroup
\allowdisplaybreaks
\begin{align}
\label{eq:channel_gg}
\mathcal{H}^{(L)}_{\mathbf{gg}} & = \sumhel \big[A^{(0)}(1_{\bar b},2_{b},3_{g},4_{g} ,5_{H})\big]^* F^{(L)}(1_{\bar b},2_{b},3_{g},4_{g} ,5_{H})  \nn
                                & \quad + \sumhel \big[A^{(0)}(1_{\bar b},2_{b},4_{g},3_{g} ,5_{H})\big]^* F^{(L)}(1_{\bar b},2_{b},4_{g},3_{g} ,5_{H}) \,, \\
\label{eq:channel_qQ}
\mathcal{H}^{(L)}_{\mathbf{q\bar{q}}} & = \sumhel \big[A^{(0)}(1_{\bar b},2_{b},3_{\bar{q}},4_{q} ,5_{H})\big]^* F^{(L)}(1_{\bar b},2_{b},3_{\bar{q}},4_{q} ,5_{H}) \,, \\
\label{eq:channel_Qq}
\mathcal{H}^{(L)}_{\mathbf{\bar{q}q}} & = \sumhel \big[A^{(0)}(1_{\bar b},2_{b},4_{\bar{q}},3_{q} ,5_{H})\big]^* F^{(L)}(1_{\bar b},2_{b},4_{\bar{q}},3_{q} ,5_{H}) \,, \\
\label{eq:channel_bB}
\mathcal{H}^{(L)}_{\mathbf{b\bar{b}}} & = \sumhel \big[ A^{(0)}(1_{\bar b},2_{b},3_{\bar{q}},4_{q} ,5_{H}) + A^{(0)}(3_{\bar b},4_{b},1_{\bar{q}},2_{q} ,5_{H}) \big]^*  \nn
                                      &   \hspace{1.5cm} \times  \big[ F^{(L)}(1_{\bar b},2_{b},3_{\bar{q}},4_{q} ,5_{H}) + F^{(L)}(3_{\bar b},4_{b},1_{\bar{q}},2_{q} ,5_{H}) \big] \nn
                                      &  + \sumhel  \big[ A^{(0)}(1_{\bar b},4_{b},3_{\bar{q}},2_{q} ,5_{H}) + A^{(0)}(3_{\bar b},2_{b},1_{\bar{q}},4_{q} ,5_{H}) \big]^*  \nn
                                      &   \hspace{1.5cm} \times  \big[ F^{(L)}(1_{\bar b},4_{b},3_{\bar{q}},2_{q} ,5_{H}) + F^{(L)}(3_{\bar b},2_{b},1_{\bar{q}},4_{q} ,5_{H}) \big] \,, \\
\label{eq:channel_Bb}
\mathcal{H}^{(L)}_{\mathbf{\bar{b}b}} & = \sumhel       \big[ A^{(0)}(1_{\bar b},2_{b},4_{\bar{q}},3_{q} ,5_{H}) + A^{(0)}(4_{\bar b},3_{b},1_{\bar{q}},2_{q} ,5_{H}) \big]^*  \nn
                                      &   \hspace{1.5cm} \times  \big[ F^{(L)}(1_{\bar b},2_{b},4_{\bar{q}},3_{q} ,5_{H}) + F^{(L)}(4_{\bar b},3_{b},1_{\bar{q}},2_{q} ,5_{H}) \big] \nn
                                      &  + \sumhel    \big[ A^{(0)}(1_{\bar b},3_{b},4_{\bar{q}},2_{q} ,5_{H}) + A^{(0)}(4_{\bar b},2_{b},1_{\bar{q}},3_{q} ,5_{H}) \big]^*  \nn
                                      &   \hspace{1.5cm} \times  \big[ F^{(L)}(1_{\bar b},3_{b},4_{\bar{q}},2_{q} ,5_{H}) + F^{(L)}(4_{\bar b},2_{b},1_{\bar{q}},3_{q} ,5_{H}) \big] \,, \\
\label{eq:channel_bb}
\mathcal{H}^{(L)}_{\mathbf{bb}} & =  \sumhel \big[ A^{(0)}(3_{\bar b},1_{b},4_{\bar{q}},2_{q} ,5_{H}) + A^{(0)}(4_{\bar b},2_{b},3_{\bar{q}},1_{q} ,5_{H}) \big]^*  \nn
                                &   \hspace{1.5cm} \times  \big[ F^{(L)}(3_{\bar b},1_{b},4_{\bar{q}},2_{q} ,5_{H}) + F^{(L)}(4_{\bar b},2_{b},3_{\bar{q}},1_{q} ,5_{H}) \big] \nn
                                &   +\sumhel \big[ A^{(0)}(3_{\bar b},2_{b},4_{\bar{q}},1_{q} ,5_{H}) + A^{(0)}(4_{\bar b},1_{b},3_{\bar{q}},2_{q} ,5_{H}) \big]^*  \nn
                                &   \hspace{1.5cm} \times  \big[ F^{(L)}(3_{\bar b},2_{b},4_{\bar{q}},1_{q} ,5_{H}) + F^{(L)}(4_{\bar b},1_{b},3_{\bar{q}},2_{q} ,5_{H}) \big] \,, \\
\label{eq:channel_BB}
\mathcal{H}^{(L)}_{\mathbf{\bar{b}\bar{b}}} & = \sumhel   \big[ A^{(0)}(1_{\bar b},3_{b},2_{\bar{q}},4_{q} ,5_{H}) + A^{(0)}(2_{\bar b},4_{b},1_{\bar{q}},3_{q} ,5_{H}) \big]^*  \nn
                                            &   \hspace{1.5cm} \times  \big[ F^{(L)}(1_{\bar b},3_{b},2_{\bar{q}},4_{q} ,5_{H}) + F^{(L)}(2_{\bar b},4_{b},1_{\bar{q}},3_{q} ,5_{H}) \big] \nn
                                            &  + \sumhel     \big[ A^{(0)}(1_{\bar b},4_{b},3_{\bar{q}},2_{q} ,5_{H}) + A^{(0)}(2_{\bar b},3_{b},1_{\bar{q}},4_{q} ,5_{H}) \big]^*  \nn
                                            &   \hspace{1.5cm} \times  \big[ F^{(L)}(1_{\bar b},4_{b},3_{\bar{q}},2_{q} ,5_{H}) + F^{(L)}(2_{\bar b},3_{b},1_{\bar{q}},4_{q} ,5_{H}) \big] \,.
\end{align}
\endgroup
We evaluate the permutations of the finite remainders numerically. The analytic continuation is performed by adding a small positive imaginary part to each $s_{ij}$ and to $p_5^2$, which is done automatically by \textsc{DiffExp}.

To facilitate future checks, we provide here the benchmark evaluation at a physical point corresponding to the $pp \to b\bar{b}H$ scattering process,
\begin{align}
&s_{12}=\frac{49}{576}\,, \quad 
s_{23}=-\frac{15337}{2048}\,, \quad 
s_{34} = \frac{63}{4}\,,\quad  
s_{45} = -\frac{288491}{38912}\,,\nonumber \\[10pt]
& \qquad \quad s_{15} = \frac{455}{64}\,, \quad 
p_5^2 = 7\,, \quad 
\mathrm{tr}_5 = i\frac{49\sqrt{50998583}}{1400832}  \,,
\label{eq:physicalpointHbb}
\end{align}
which corresponds to the following values of the momentum twistor variables,
\begin{align}
&
x_{1}=\frac{49}{576}\,, \quad 
x_{2}=-\frac{77405}{76608}-i\frac{\sqrt{50998583}}{76608}\,, \quad
x_{3}=-\frac{557251}{411874}+i\frac{95\sqrt{50998583}}{411874}\,, \quad \nonumber \\[10pt]
& \qquad \quad
x_{4}=-\frac{2817}{32}\,,\quad 
x_{5}=-\frac{11290}{41629} - i\frac{2\sqrt{50998583}}{41629}\,, \quad 
x_{6}=-\frac{370917}{4256} \,.
\label{eq:physicalpointHbbMomTwistor}
\end{align}
The values of the bare two-loop amplitudes normalised by the tree-level amplitudes,
$\hat{A}^{(L)} = A^{(L)}/A^{(0)}$,
for the independent mostly-plus helicity configurations of the subprocesses $0\to \bar{b} b gg H$ and $0\to \bar{b} b \bar{q} q H$ are given in Tables~\ref{tab:benchmark2g2bHbare} and~\ref{tab:benchmark2q2bHbare}. Table~\ref{tab:benchmarkfinremsq2L} shows the values of the two-loop reduced squared finite remainders $\mathcal{H}^{(2)}$. 
Analogous tables for the one-loop amplitudes are given in Appendix~\ref{app:oneloop}.
\begin{table}[t!]
\centering
\begin{tabularx}{1.0\textwidth}{|C{0.7}|C{0.8}|C{0.6}|C{1.2}|C{1.2}|C{1.2}|C{1.3}|}
\hline
 $\bbggh$     & helicity & $\eps^{-4}$ & $\eps^{-3}$ & $\eps^{-2}$ & $\eps^{-1}$ & $\eps^{0}$ \\
\hline
$\hat A^{(2),1}$ & $++++$ & $ 4.5$ & $ -11.9857 + 9.42478 i$ & $ 1.38005 - 40.6951 i$ & $ 37.5629 + 74.9878 i$ & $ -160.364 - 16.4633 i $ \\
                 & $+++-$ & $ 4.5$ & $ -11.9857 + 9.42478 i$ & $ 11.3257 - 12.3672 i$ & $ -26.8161 + 82.1522 i$ & $ -142.327 - 160.925 i $ \\
                 & $++-+$ & $ 4.5$ & $ -11.9857 + 9.42478 i$ & $ 2.69154 - 41.4561 i$ & $ 35.9446 + 68.3748 i$ & $ -132.233 - 11.7912 i $ \\
                 & $++--$ & $ 4.5$ & $ -11.9857 + 9.42478 i$ & $ 21.8803 - 71.2779 i$ & $ 85.0932 + 67.5004 i$ & $ -293.742 + 11.2118 i $ \\
\hline
$\hat A^{(2),n_f}$  & $++++$ & $ 0$ & $ 0.5$ & $ -0.177826 + 1.04486 i$ & $ -0.769158 - 3.80277 i$ & $ -5.39544 + 7.05528 i $ \\
                    & $+++-$ & $ 0$ & $ 0.5$ & $ -0.192856 + 1.0472 i$ & $ 1.4513 + 2.42621 i$ & $ -3.57357 + 44.5555 i $ \\
                    & $++-+$ & $ 0$ & $ 0.5$ & $ -0.192856 + 1.0472 i$ & $ -0.467396 - 4.03798 i$ & $ -3.83854 + 2.69906 i $ \\
                    & $++--$ & $ 0$ & $ 0.5$ & $ 0.987798 + 0.631652 i$ & $ 3.957 - 5.16329 i$ & $ 33.7155 - 38.6759 i $ \\
\hline
$\hat A^{(2),n_f^2}$ & $++++$ & $ 0$ & $ 0$ & $ 0$ & $ 0.00334 - 0.000519914 i$ & $ 0.00266436 + 0.0210796 i $ \\
                     & $+++-$ & $ 0$ & $ 0$ & $ 0$ & $ 0$ & $ 0 $ \\
                     & $++-+$ & $ 0$ & $ 0$ & $ 0$ & $ 0$ & $ 0 $ \\
                     & $++--$ & $ 0$ & $ 0$ & $ 0$ & $ 0.262368 - 0.0923434 i$ & $ 0.532893 + 1.66516 i $ \\
\hline
\end{tabularx}
\caption{\label{tab:benchmark2g2bHbare} Numerical values of the bare $\bbggh$ partial amplitudes at two loops (normalised to the tree-level amplitude) at the kinematic point in Eq.~\eqref{eq:physicalpointHbbMomTwistor} for the four independent helicity configurations and the various closed fermion loops contributions.}
\end{table}

\begin{table}[t!]
\centering
\begin{tabularx}{1.0\textwidth}{|C{0.7}|C{0.8}|C{0.6}|C{1.2}|C{1.2}|C{1.2}|C{1.3}|}
\hline
 $\bbqqh$     & helicity & $\eps^{-4}$ & $\eps^{-3}$ & $\eps^{-2}$ & $\eps^{-1}$ & $\eps^{0}$ \\
\hline
$\hat A^{(2),1}$ & $+++-$ & $ 2 $ & $ -6.81012$ & $ 25.5694 + 17.9036 i$ & $ -60.3404 - 6.4188 i$ & $ 48.2991 - 125.381 i $ \\
                 & $++-+$ & $ 2 $ & $ -6.81012$ & $ 22.4573 + 14.9001 i$ & $ -60.7797 + 3.42105 i$ & $ 96.4449 - 180.941 i $ \\
\hline
$\hat A^{(2),n_f}$ & $+++-$ & $ 0$ & $ 1.66667$ & $ -4.60863 + 4.18879 i$ & $ 13.2979 - 5.52188 i$ & $ 4.96804 + 95.7191 i $ \\
                   & $++-+$ & $ 0$ & $ 1.66667$ & $ -4.60863 + 4.18879 i$ & $ 11.2232 - 7.52422 i$ & $ -1.06892 + 93.2862 i $ \\
\hline
$\hat A^{(2),n_f^2}$ & $+++-$ & $ 0$ & $ 0$ & $ 0.444444$ & $ -0.969043 + 2.79253 i$ & $ -6.91677 - 6.08868 i $ \\
                     & $++-+$ & $ 0$ & $ 0$ & $ 0.444444$ & $ -0.969043 + 2.79253 i$ & $ -6.91677 - 6.08868 i $ \\
\hline
\end{tabularx}
\caption{\label{tab:benchmark2q2bHbare} Numerical values of the bare $\bbqqh$ partial amplitudes at two loops (normalised to the tree level amplitude) at the kinematic point in Eq.~\eqref{eq:physicalpointHbbMomTwistor} for the four independent helicity configurations and the various closed fermion loops contributions. }
\end{table}

\begin{table}[t!]
\centering
\begin{tabularx}{0.85\textwidth}{|C{0.7}|C{1.1}|C{1.1}|C{1.1}|}
\hline
channel      & $\mathrm{Re}\;\mathcal{H}^{(2),1}$ & $\mathrm{Re}\;\mathcal{H}^{(2),n_f}$ & $\mathrm{Re}\;\mathcal{H}^{(2),n_f^2}$ \\
\hline
$\mathbf{gg}$         & $ 156680.6267$ & $ -41215.80337$ & $ 405.9379563 $ \\
$\mathbf{q\bar{q}}$   & $ 0.09391314268$ & $ -0.02045942258$ & $ -0.004225713438 $ \\
$\mathbf{\bar{q}q}$   & $ 0.3494872243$ & $ -0.08069122736$ & $ -0.004225713438 $ \\
$\mathbf{b\bar{b}}$   & $ 48640.80398$ & $ -26530.01855$ & $ 2458.442153 $ \\
$\mathbf{\bar{b}b}$   & $ -141130.5373$ & $ 42183.03094$ & $ 3711.445449 $ \\
$\mathbf{bb}/\mathbf{\bar{b}\bar{b}}$  & $ -53679.25708$ & $ 1988.662899$ & $ 894.7895467 $ \\
\hline
\end{tabularx}
\caption{\label{tab:benchmarkfinremsq2L} Numerical values of the two-loop reduced squared finite
remainders $\mathcal{H}^{(2)}$ defined in Eqs.~\eqref{eq:channel_gg}-\eqref{eq:channel_BB} at the kinematic point in Eq.~\eqref{eq:physicalpointHbbMomTwistor} for the various closed fermion loops contributions and the scattering channels specified in Eq.~\eqref{eq:channel_definition}.}
\end{table}

\begin{figure}[t!]
\centering
\begin{subfigure}{.5\textwidth}
    \centering
    \includegraphics[width=.9\textwidth]{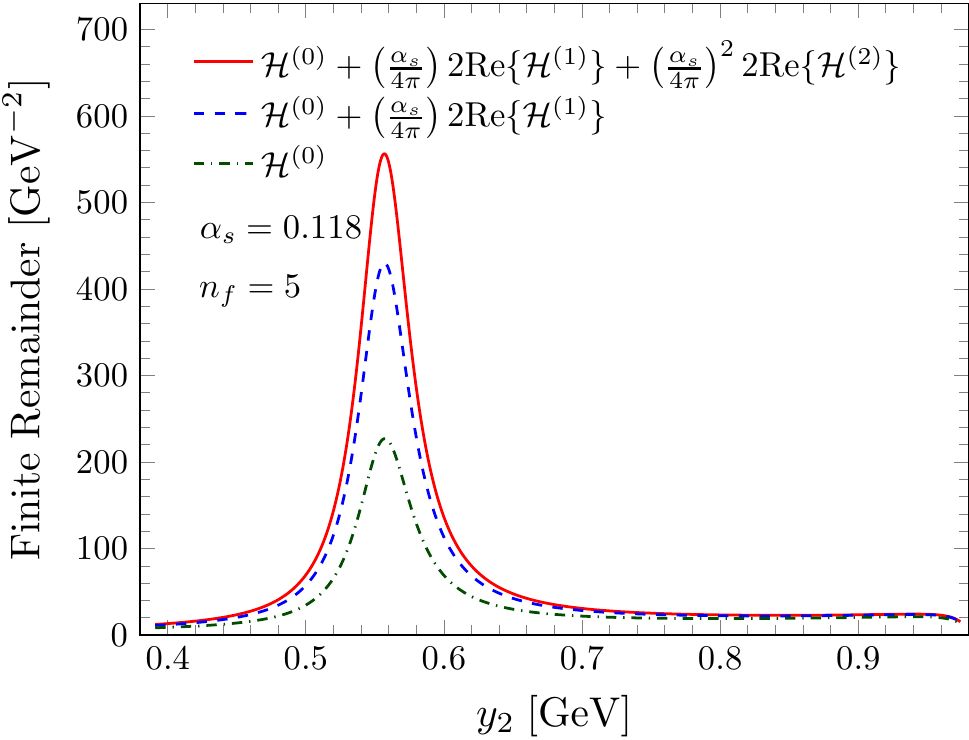}
    \label{fig:gg}
    \caption{$\mathbf{gg}$}
\end{subfigure}%
\begin{subfigure}{.5\textwidth}
    \centering
    \includegraphics[width=.9\textwidth]{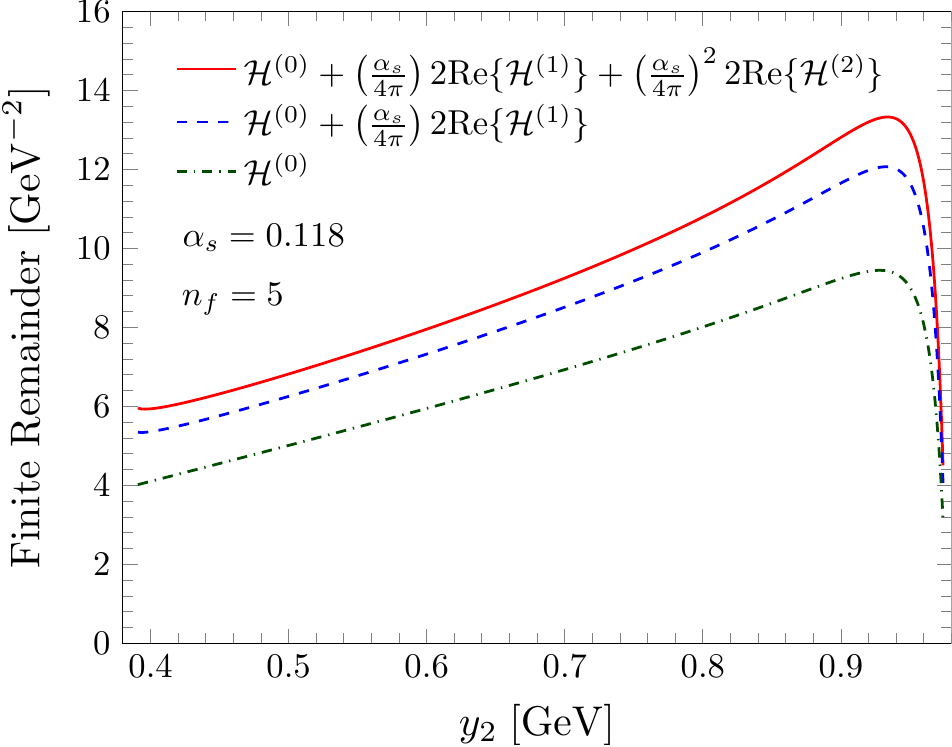}
    \label{fig:qqx}
    \caption{$\mathbf{q\bar{q}}$}
\end{subfigure}
\begin{subfigure}{.5\textwidth}
    \centering
    \includegraphics[width=.9\textwidth]{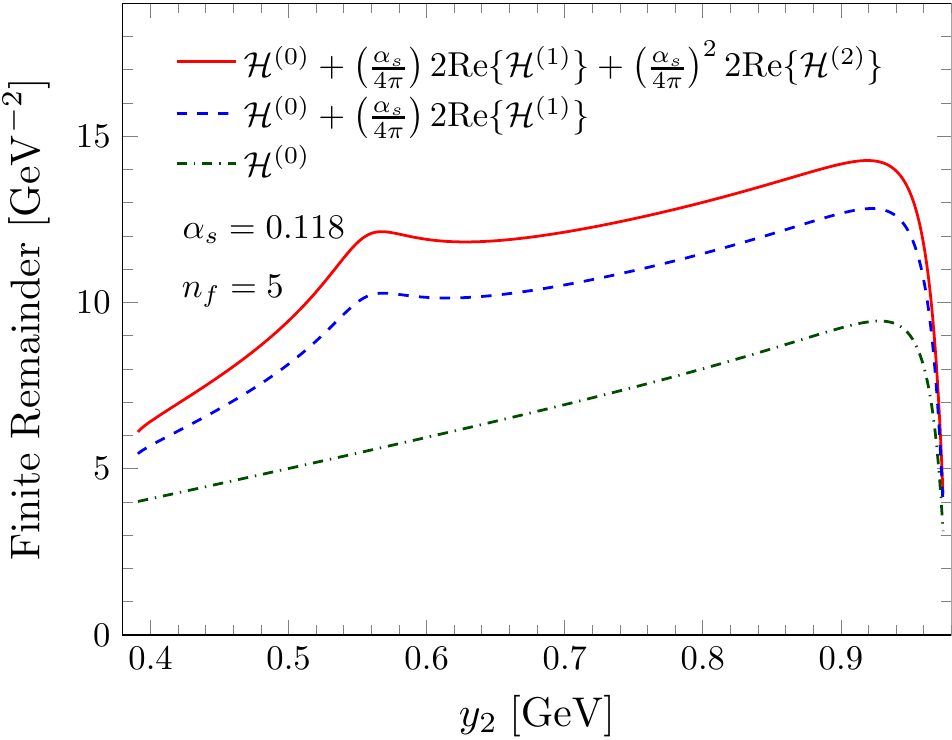}
    \label{fig:qxq}
    \caption{$\mathbf{\bar{q}q}$}
\end{subfigure}%
\begin{subfigure}{.5\textwidth}
    \centering
    \includegraphics[width=.9\textwidth]{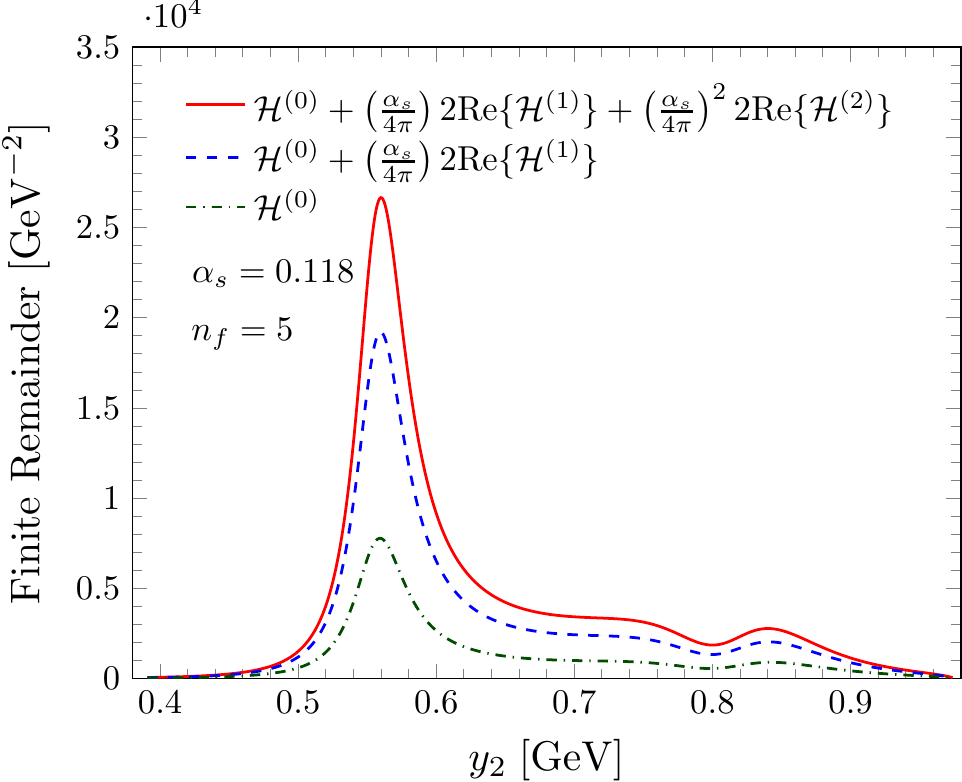}
    \label{fig:bb}
    \caption{$\mathbf{bb}$}
\end{subfigure}
\begin{subfigure}{.5\textwidth}
    \centering
    \includegraphics[width=.9\textwidth]{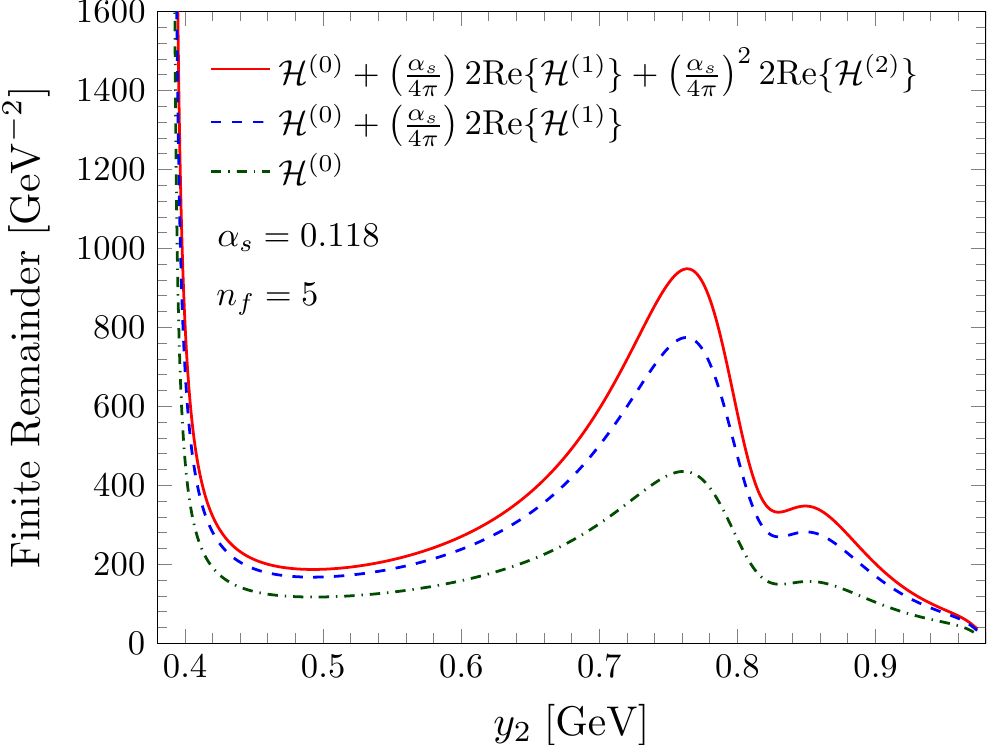}
    \label{fig:bbx}
    \caption{$\mathbf{b\bar{b}}$}
\end{subfigure}%
\begin{subfigure}{.5\textwidth}
    \centering
    \includegraphics[width=.9\textwidth]{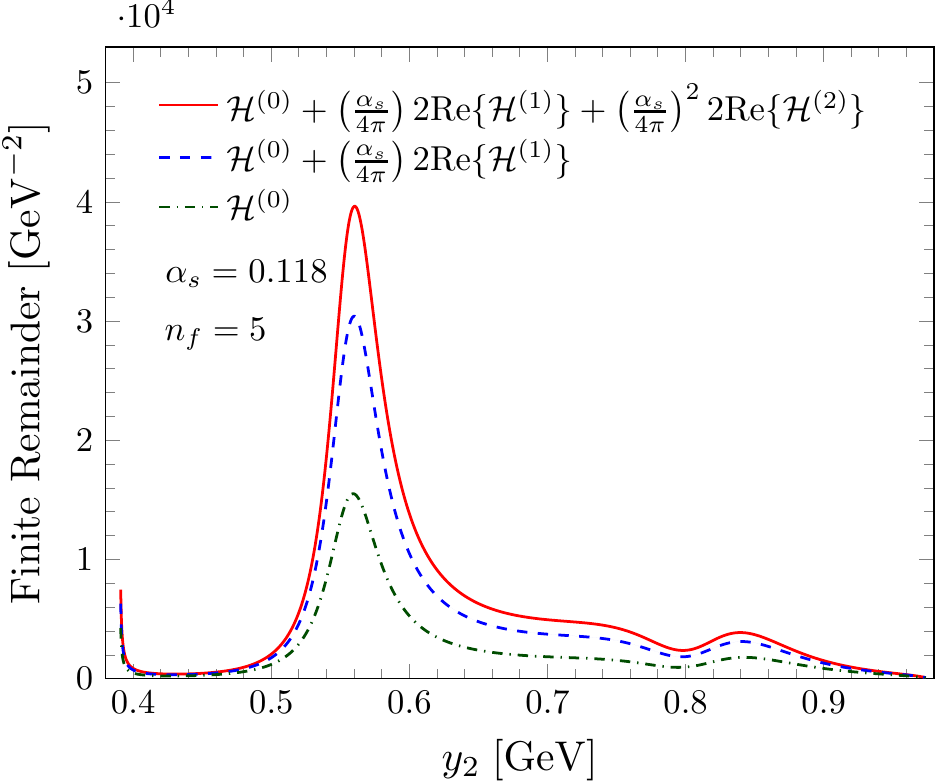}
    \label{fig:bxb}
    \caption{$\mathbf{\bar{b}b}$}
\end{subfigure}
\caption[short]{Reduced squared finite remainders $\mathcal{H}^{(L)}$ at tree level, one and two loops evaluated on the one-dimensional phase space slice defined 
in Eq.~\eqref{eq:parametrisation}, as functions of the variable $y_2$, for the channels defined in Eq.~\eqref{eq:channel_definition}.}
\label{fig:plots}
\end{figure}

To demonstrate the suitability of our results for phenomenological applications we present the evaluation of the finite remainders interfered with the tree-level amplitudes on a univariate slice of the phase space. For this purpose we parametrise the momenta for the scattering processes~\eqref{eq:subprocesses} in terms of angles and energy fractions of the final state as
\begin{equation} \label{eq:parametrisation}
\begin{aligned}
& p_1 = \frac{y_1 \sqrt{s}}{2} \left( 1\,, 1\,, 0\,, 0 \right) \,, \quad \\
& p_2 = \frac{y_2 \sqrt{s}}{2} \left( 1\,, \cos\theta\,, -\sin\theta \sin\phi \,, -\sin\theta \cos\phi \right) \,,  \\
& p_3 = \frac{\sqrt{s}}{2} \left(-1\,,0\,, 0 \,, -1 \right) \,,  \\
& p_4 = \frac{\sqrt{s}}{2} \left(-1\,,0\,, 0 \,, 1 \right) \,,  \\
\end{aligned}
\end{equation}
while $p_5$ is fixed by momentum conservation, $p_5 = -p_1-p_2-p_3-p_4$. We have chosen the particle with momentum $p_1$ to be produced at an angle of $\pi/2$ with respect to the beam axis.
Requiring that the Higgs boson is on-shell, $p_5^2 = m_H^2$, constrains the angle $\theta$ as
\begin{align}
\cos\theta = 1+\frac{2}{y_1 y_2}\left(1-y_1-y_2-\frac{m_H^2}{s} \right)\,.
\end{align}
To restrict the kinematics to a one-dimensional slice we choose
\begin{align} \label{eq:fixparameters}
s = 1 \,, \quad \qquad \phi = \frac{1}{10} \,, \quad \qquad y_1 = \frac{3}{5} \,, \quad \qquad m_H = \frac{1}{10} \,.
\end{align}
The reality of the angle $\theta$ then restricts the free parameter $y_2$ to the interval $y_2 \in
[\frac{39}{100},\frac{39}{40}]$. In order to evaluate the finite remainders for all the processes
shown in Eq.~\eqref{eq:channel_definition} we need to compute the finite remainder special functions
at $16$ permutations of each phase space point, as can be seen explicitly in
Eqs.\eqref{eq:channel_gg}~-~\eqref{eq:channel_BB}. We do this by integrating the system of canonical differential equations~\eqref{eq:dh} with \textsc{DiffExp} using rationalised values of the invariants. For each permutation of each point we compute the ``distance'' (technically speaking, the number of segments into which the integration contour is divided by \textsc{DiffExp}) from the six reference points provided in the ancillary files, and choose the closest one as initial point for the integration. In Figure~\ref{fig:plots} we plot the values of the one- and two-loop reduced squared finite remainders for all the processes defined in Eq.~\eqref{eq:channel_definition}, as functions of the parameter $y_2$.

We stress that the purpose of the plots in Figure~\ref{fig:plots} is merely to demonstrate that our results for the finite remainders can be evaluated reliably in the physical scattering region. Nothing can be inferred about the convergence of the perturbative expansion at the cross section level. One interesting feature which can be appreciated from Figure~\ref{fig:plots} is the appearance of a loop-induced peak in the finite remainders for the channel $\mathbf{\bar{q}q}$. The peak is absent at tree level for the same channel and up to two loops for $\mathbf{q\bar{q}}$. The latter channel is related to $\mathbf{\bar{q}q}$ by the exchange $3\leftrightarrow 4$ of the external particles. We observe that the peak stems from the values of the finite remainder function basis, while the rational coefficients are not enhanced in that region. In order to pinpoint more precisely the origin of this phenomenon, we construct the explicit analytic expressions for some of the special functions which exhibit the peak, starting from their iterated integral expression obtained by solving the system of canonical differential equations~\eqref{eq:dh}. For instance,
\begin{equation} \label{eq:h24}
\begin{aligned}
h^{(2)}_4 & = 2 \, \text{Li}_2\left( 1 - \frac{s_{15}}{p_5^2} \right) + 2 \, \text{Li}_2 \left( 1- \frac{s_{23}}{s_{15}}\right) -\frac{\pi^2}{4} + \frac{1}{2} \log^2\left( p_5^2 \right) + \frac{1}{2}\log^2\left( -s_{45}\right) + 2 \log^2\left( s_{15}\right) \\
& - \frac{1}{2}\log^2\left( -s_{23}\right) - 2 \log\left( s_{15}\right) \log\left( s_{15} - s_{23}\right) + \log^2\left( s_{15} - s_{23}\right) + \log\left( p_5^2\right) \log\left(-s_{23}\right)  \\
& - 2 \log\left( p_5^2\right) \log\left(s_{15} \right)  - \log\left( p_5^2\right) \log\left( -s_{14}\right) +2 \log\left( s_{15}\right) \log\left( -s_{14}\right) 
 - \log\left(-s_{23}\right) \log\left( -s_{14}\right) \\
 & +i \pi \biggl[ \log\left( p_5^2\right) - 2 \log\left( s_{15}\right) + 2 \log\left( s_{15} - s_{23}\right) - \log\left( -s_{23}\right) - \log\left( -s_{14}\right)\biggr] \,,
\end{aligned}
\end{equation}
which is well defined in the $s_{34}$ physical scattering region. The analytic continuation to any other region is obtained by adding a small positive imaginary part to each $s_{ij}$ and to $p_5^2$. We checked that the values of $h^{(2)}_4$ as given by Eq.~\eqref{eq:h24} (and of its permutations) agree with the evaluation through the generalised series expansion. The function $h^{(2)}_4$ exhibits no peak on the phase space sliced defined by Eqs.~\eqref{eq:parametrisation} and~\eqref{eq:fixparameters}, and indeed the finite remainder for the channel $\mathbf{q\bar{q}}$ does not exhibit such feature. The permutation $3\leftrightarrow 4$ of $h^{(2)}_4$, which contributes to $\mathbf{\bar{q}q}$, is instead peaked around $y_2 \approx 0.5566$. Thanks to the analytic expression~\eqref{eq:h24} we can identify the source of the peak in the logarithms of $s_{24}$, which originate from the $\log (-s_{23})$'s in Eq.~\eqref{eq:h24} upon swapping $3\leftrightarrow 4$. Indeed the tree-level amplitude for the subprocess $0\to\bar{b} b \bar{q} q H$, given by Eq.~\eqref{eq:A0bbqqh}, is manifestly free of $1/\langle 23 \rangle$ poles. The $0\to\bar{b} b \bar{q} q H$ diagrams with a $1/s_{23}$ pole would come with a loop and so end up scaling as $\log(-s_{23})$. While $\log(-s_{23})$ is not enhanced on the one-dimensional slice under consideration, its $3\leftrightarrow4$ permutation $\log(-s_{24})$, which contributes to $\mathcal{H}^{(L)}_{\mathbf{\bar{q}q}}$ for $L=1,2$, is peaked at $y_2 \approx 0.5566$, where $s_{24}$ reaches its minimum absolute value on the slice. 
The tree-level amplitudes for the channel $\mathbf{gg}$ instead do have poles at $s_{23}=0$, which
can be seen explicitly in Eqs.~\eqref{eq:A0bbggh}. Since $\mathcal{H}^{(L)}_{\mathbf{gg}}$ with
$L=1,2$ receive contributions from the partial finite remainders in both the standard orientation
and with the swap $3\leftrightarrow4$, as shown in Eq.~\eqref{eq:channel_gg}, their plot in Figure~\ref{fig:plots}~(a) exhibits this peak already at tree level. The same holds for the $\mathbf{bb}$ and $\mathbf{\bar{b}b}$ channels, as can be seen in Figures~\ref{fig:plots}~(d) and~(f). 

Also in Figure~\ref{fig:plots}, we observe divergences at $y_2 = 39/100$ for the processes
$\mathbf{b\bar{b}}$ and $\mathbf{\bar{b}b}$. This divergence is associated with the propagator
$1/s_{12}$, which can only appear in processes with two pairs of bottom quarks. In Eqs.~\eqref{eq:channel_bB}~and~\eqref{eq:channel_Bb} 
we can see the $q\bar{q}$ fermion pairs can appear
with momenta $p_1$ and $p_2$, which is not the case for other processes. All the other features of the plots
in Figure~\ref{fig:plots} can be similarly understood in terms of tree-level propagators going on shell.

\section{Conclusions}
\label{sec:conclusions}

In this article we have presented an analytic form for the two-loop QCD corrections to the process
$pp\to b\bar{b}H$. To the best of our knowledge this is the first complete set of helicity
amplitudes provided for a $2\to 3$ scattering process with an off-shell leg. A special function basis for
the finite remainder was identified obeying a canonical form differential equation. The method for
constructing such a basis was presented in the recent $pp\to Wb\bar{b}$ computation~\cite{Badger:2021nhg} by three of the present
authors. The finite remainders were then extracted from
multiple evaluations over finite fields using a rational phase-space and IBP reduction. We obtained relatively compact
results after determining the linear relations between the rational coefficients and
performing a univariate partial fraction decomposition on the fly. The final expressions were evaluated using the method
of generalised series expansions as implemented in the \textsc{DiffExp} code~\cite{Hidding:2020ytt}.

The expressions have been validated in a number of ways and we observe that they exhibit a smooth behaviour in all
scattering regions. Evaluation times appear to be suitable for phenomenological applications,
especially since we have not tried to optimise the route through the phase-space evaluations as has
been done in other applications of the method~\cite{Frellesvig:2019byn,Abreu:2020jxa,Becchetti:2020wof,Bonciani:2021zzf,abreu2021twoloop}.

The techniques presented here show promise for applications to other important scattering processes such
as $pp\to V+2j$ and $pp\to H+2j$, although the non-planar sector is not entirely known and remains a high priority.

The analytic expressions together with \texttt{Mathematica} scripts to evaluate them numerically are provided in the ancillary files accompanying this article.

\acknowledgments
SZ wishes to thank Dmitry Chicherin and Vasily Sotnikov for useful discussions.
This project received funding from the European Union's Horizon 2020 research and innovation programmes
\textit{New level of theoretical precision for LHC Run 2 and beyond} (grant agreement No 683211),
\textit{High precision multi-jet dynamics at the LHC} (grant agreement No 772009), and
\textit{Novel structures in scattering amplitudes} (grant agreement No 725110).
HBH was partially supported by STFC consolidated HEP theory grant ST/T000694/1.
SZ gratefully acknowledges the computing resources provided by the Max Planck Institute for Physics and by the Max Planck Computing \& Data Facility.

\appendix

\section{Renormalisation Constants}
\label{app:renormconstants}

The one- and two-loop bottom-quark Yukawa renormalisation constants entering the UV counterterm in Eqs.~\eqref{eq:poles1L}~and~\eqref{eq:poles2L} are
\begin{align}
s_1 & = -\frac{3 C_F}{2\eps} \,, \\
s_2 & = \frac{3}{8\eps^2}\big(3 C_F^2 + \beta_0 C_F \big) - \frac{1}{8\eps} \bigg( \frac{3 C_F^2}{2} + \frac{97}{6} C_F C_A -\frac{10}{3} C_F T_F n_f   \bigg) \,,
\end{align}
while the $\beta$ function coefficients and anomalous dimensions are

\begingroup
\allowdisplaybreaks
\begin{align}
\beta_0 = & \;\frac{11}{3}C_A - \frac{4}{3} T_F n_f \,, \\
\beta_1 = & \; \frac{34}{3} C_A^2 - \frac{20}{3} C_A T_F n_f - 4 C_F T_F n_f \,, \\
\gamma_0^g = & \; -\frac{11}{3}C_A + \frac{4}{3} T_F n_f \,,   \\
\gamma_1^g = & \;  C_A^2 \left( -\frac{692}{27} + \frac{11\pi^2}{18} + 2 \zeta_3\right)
               + 4 C_F T_F n_f
               + C_A T_F n_f \left( \frac{256}{27} - \frac{2\pi^2}{9}\right) \,, \\
\gamma_0^q = & \; -3 C_F \,, \\
\gamma_1^q = & \; C_F^2 \left( -\frac{3}{2} + 2 \pi^2 - 24 \zeta_3 \right)
                  + C_F C_A \left( -\frac{961}{54} -\frac{11\pi^2}{6} + 26 \zeta_3 \right) \nn
             & \;  + C_F T_F n_f \left( \frac{130}{27} + \frac{2\pi^2}{3} \right) \,, \\
\gamma_0^\cusp = & \; 4 \,, \\
\gamma_1^\cusp = & \; \left( \frac{268}{9} - \frac{4\pi^2}{3} \right) C_A -\frac{80}{9} T_F n_f \,,
\end{align}
\endgroup
where $C_A = N_c$ and $C_F = (N_c^2-1)/(2N_c)$.

\section{One-Loop Results}
\label{app:oneloop}

\begin{table}[t!]
\centering
\begin{tabularx}{1.0\textwidth}{|C{0.7}|C{0.8}|C{0.5}|C{1.2}|C{1.2}|C{1.3}|C{1.3}|}
\hline
 $\bbggh$     & helicity & $\eps^{-2}$ & $\eps^{-1}$ & $\eps^{0}$ & $\eps^{1}$ & $\eps^{2}$ \\
\hline
$\hat A^{(1),1}$ & $++++$ & $ -3 $ & $ 3.07857 - 3.14159 i$ & $ 0.317351 + 8.42128 i$ & $ -1.25257 - 8.56907 i$ & $ 25.8294 - 4.35648 i $ \\
                 & $+++-$ & $ -3 $ & $ 3.07857 - 3.14159 i$ & $ -2.99786 - 1.02133 i$ & $ 2.86487 - 28.7164 i$ & $ 30.3093 - 26.3373 i $ \\
                 & $++-+$ & $ -3 $ & $ 3.07857 - 3.14159 i$ & $ -0.119814 + 8.67497 i$ & $ -1.43041 - 5.33656 i$ & $ 19.6373 - 0.110475 i $ \\
                 & $++--$ & $ -3 $ & $ 3.07857 - 3.14159 i$ & $ -6.51606 + 18.6156 i$ & $ -21.7849 + 24.0036 i$ & $ -24.6605 + 55.6878 i $ \\
\hline
$\hat A^{(1),n_f}$ & $++++$ & $ 0$ & $ 0$ & $ -0.005010 + 0.000779871 i$ & $ -0.00700827 - 0.0150298 i$ & $ 0.0109029 - 0.0163643 i $ \\
                   & $+++-$ & $ 0$ & $ 0$ & $ 0$ & $ 0$ & $ 0 $ \\
                   & $++-+$ & $ 0$ & $ 0$ & $ 0$ & $ 0$ & $ 0 $ \\
                   & $++--$ & $ 0$ & $ 0$ & $ -0.393552 + 0.138515 i$ & $ -0.793221 - 1.11035 i$ & $ 0.635641 - 1.48796 i $ \\
\hline
$\bbqqh$     & helicity & $\eps^{-2}$ & $\eps^{-1}$ & $\eps^{0}$ & $\eps^{1}$ & $\eps^{2}$ \\
\hline
$\hat A^{(1),1}$ & $+++-$ & $ -2 $ & $ 2.48840$ & $ -9.99430 - 8.95182 i$ & $ 2.20899 - 24.3401 i$ & $ 4.76962 - 27.6604 i $ \\
                 & $++-+$ & $ -2 $ & $ 2.48840$ & $ -8.43825 - 7.45006 i$ & $ 7.21741 - 24.6383 i$ & $ 13.6369 - 20.4876 i $ \\
\hline
$\hat A^{(1),n_f}$ & $+++-$ & $ 0$ & $ -0.666667$ & $ 0.726782 - 2.09440 i$ & $ 2.29387 + 2.28325 i$ & $ -2.54017 + 0.316127 i $ \\
                   & $++-+$ & $ 0$ & $ -0.666667$ & $ 0.726782 - 2.09440 i$ & $ 2.29387 + 2.28325 i$ & $ -2.54017 + 0.316127 i $ \\
\hline
\end{tabularx}
\caption{\label{tab:benchmarkbare1L} Numerical values of the bare $\bbggh$ and $\bbqqh$ partial amplitudes at one loop (normalised to the tree-level amplitude) at the kinematic point in 
Eq.~\eqref{eq:physicalpointHbbMomTwistor} for the four independent helicity configurations and the various closed fermion loops contributions. }
\end{table}

\begin{table}[t!]
\centering
\begin{tabularx}{0.85\textwidth}{|C{0.7}|C{1.1}|C{1.1}|C{1.1}|}
\hline
channel & $\mathcal{H}^{(0)}$ & $\mathrm{Re}\;\mathcal{H}^{(1),1}$ & $\mathrm{Re}\;\mathcal{H}^{(1),n_f}$  \\
\hline
$\mathbf{gg}$       & $ 1121.375369 $    & $ 4905.689964$ & $ 204.1069797 $ \\
$\mathbf{q\bar{q}}$ & $ 0.001095232986 $ & $ -0.008958148524$ & $ 0.0007959961305 $ \\
$\mathbf{\bar{q}q}$ & $ 0.001095232986 $ & $ 0.01182947634$ & $ 0.0007959961305 $ \\
$\mathbf{b\bar{b}}$ & $ 738.4111805 $ & $ 5948.275150$ & $ -2005.976183 $ \\
$\mathbf{\bar{b}b}$ & $ 774.9861507 $ & $ -8346.007933$ & $ -2253.325645 $ \\
$\mathbf{bb}/\mathbf{\bar{b}\bar{b}}$ & $ 71.81424881 $  & $ -678.1382010$ & $ -243.5040325 $ \\
\hline
\end{tabularx}
\caption{\label{tab:benchmarkfinremsq1L} Numerical values of the tree-level reduced squared amplitudes  $\mathcal{H}^{(0)}$ and 
one-loop reduced squared finite remainders $\mathcal{H}^{(1)}$ 
defined in Eqs.~\eqref{eq:channel_gg}-\eqref{eq:channel_BB} at the kinematic point in 
Eq.~\eqref{eq:physicalpointHbbMomTwistor} for  the closed fermion loops contributions and the scattering channels specified in Eq.~\eqref{eq:channel_definition}.}
\end{table}

We present in Table~\ref{tab:benchmarkbare1L} the numerical values of the one-loop bare helicity amplitudes at the
kinematic point given in Eq.~\eqref{eq:physicalpointHbbMomTwistor}, evaluated through $O(\eps^2)$ for the different closed fermion loop contributions defined in Eq.~\eqref{eq:NfDecomposition1L}. 
These one-loop amplitudes are required in the computation of the two-loop pole terms in Eqs.~\eqref{eq:poles1L}~and~\eqref{eq:poles2L}. 
Furthermore, Table~\ref{tab:benchmarkfinremsq1L} shows the tree-level reduced squared amplitudes and
one-loop reduced squared finite remainders for the scattering channels listed in Eq.~\eqref{eq:channel_definition}, at the same kinematic point.

\bibliographystyle{JHEP}
\bibliography{planar_bbh}

\end{document}